\documentclass[11pt]{article}
\usepackage{fullpage, tikz}
\usepackage[normalem]{ulem}
\usetikzlibrary{positioning}
\usepackage[T1]{fontenc}

\usepackage{geometry}
\usepackage{xcolor}
\usepackage{dirtytalk}
\usepackage{enumitem}
\usepackage{amsfonts}
\usepackage{amsthm}
\usepackage{amsmath}
\usepackage{amssymb}
\usepackage{mathtools}
\usepackage{bbm}
\usepackage[hidelinks,hypertexnames=false]{hyperref}
\usepackage{cleveref}
\usepackage{algorithm}
\usepackage[noend]{algpseudocode}

\usepackage{natbib}
\usepackage{thmtools}
\usepackage{thm-restate}
\usepackage{tikz}
\usetikzlibrary{arrows.meta,patterns,calc}

\renewcommand{\cite}[1]{\citep{#1}}
\newcommand{\textcite}[1]{\citet{#1}}

\makeatletter
\@ifundefined{newcounteralias}{}{%
  \renewcommand\thmt@autorefsetup{\@xa\def\csname\thmt@envname autorefname\@xa\endcsname\@xa{\thmt@thmname}}%
}
\makeatother

\newtheorem{theorem}{Theorem}[section]
\newtheorem{inftheorem}{Informal Theorem}

\newtheorem{claim}[theorem]{Claim}
\newtheorem{lemma}[theorem]{Lemma}

\newtheorem{proposition}[theorem]{Proposition}

\theoremstyle{definition}

\newcommand{\R}{\mathbb{R}}

\newcommand{\one}{\mathbf{1}}%

\renewcommand{\Pr}{\mathop{\mathbf{Pr}}\nolimits}

\DeclareMathOperator*{\argmin}{arg\,min}

\DeclareMathAlphabet{\mymathbb}{U}{BOONDOX-ds}{m}{n}

\newcommand{\welfare}{\mathsf{Welfare}}

\newcommand{\compratio}{\mathsf{CompRatio}}
\newcommand{\E}{\mathbb{E}}
\newcommand{\D}{\mathcal{D}}

\newcommand{\Z}{\mathbb{Z}}

\newcommand{\vals}{\mathbf{v}}

\newcommand{\hopt}{\operatorname{HOpt}}
\newcommand{\prophet}{\operatorname{Prophet}}

\newcommand{\eaopt}{\operatorname{ExAnteOpt}}

\newcommand{\Unif}{\mathrm{Unif}}

\usepackage{pgfplots}
\usepgfplotslibrary{fillbetween}
\pgfplotsset{compat=newest}

\title{How Many Prices Suffice?\\ Pricing in Multi-Unit Combinatorial Prophet Inequalities}
\author{
Zhiyi Huang \\
\small University of Texas, Austin \\
\small \texttt{zhiyih@cs.utexas.edu}
}
\date{}

\begin{document}

\allowdisplaybreaks

\maketitle
\thispagestyle{empty}
\addtocounter{page}{-1}

\begin{abstract}
We study posted-price mechanisms for social welfare maximization in multi-unit combinatorial prophet inequalities, where a seller has at least $k$ identical copies of each item and buyers with independent combinatorial valuations arrive in an unknown fixed order. In \emph{supply-based pricing}, the seller commits in advance to a nondecreasing price schedule for each item's copies using at most $r$ distinct price levels per item; the special case $r=1$ is \emph{static item pricing}. Whereas static pricing approaches the hindsight optimum as $k$ grows for a single item, in combinatorial auctions static prices are only known to guarantee $1/2$ for fractionally subadditive (XOS) valuations~\cite{feldman2014combinatorial}. At the other extreme, \textcite{ChawlaDHW25} show that unrestricted supply-based pricing ($r=k$) achieves $1-(k/(k+1))^k \to 1-1/e$ for XOS buyers, whereas fully dynamic pricing approaches $1$. How does the achievable competitive ratio depend on the supply $k$ and the number of price levels $r$?

We resolve this question across the full spectrum of price complexity, even for unit-demand buyers. First, we prove that the worst-case competitive ratio of static pricing is \emph{exactly} $1/2$ for every fixed supply $k$: unlike in the single-item setting, increasing the supply yields no worst-case improvement in the combinatorial setting. Second, we show a sharp jump from one price to two: for unit-demand buyers, a randomized mechanism using just two prices per item ($r=2$) already achieves $9/13 - o_k(1)$ (surpassing $1-1/e$) against the stronger ex-ante optimum.\footnote{For each fixed number \(r\) of price levels, \(o_k(1)\) denotes an error term that tends to zero as the minimum supply \(k\) grows, uniformly over all instances.} More generally, for every fixed $r \ge 2$, using at most $r$ prices per item achieves $\frac{1}{1+(r/(r+1))^r} - o_k(1)$, approaching $e/(e+1)$ as $r$ grows. Finally, we prove that even with unrestricted supply-based pricing (arbitrary $r$), no mechanism can guarantee a competitive ratio strictly above $2\sqrt{2}-2 \approx 0.828$ for any fixed supply $k$. Thus, supply-based pricing cannot approach $1$ as the supply grows, strictly separating it from dynamic pricing. All of our impossibility results hold even for unit-demand buyers and randomized prices.

\end{abstract}

\newpage

\section{Introduction}
\label{sec:introduction}

Consider a theater selling tickets to several screenings. Each customer wants one ticket and has a value for each screening she can attend. At the prices the theater sets for each screening, she buys the ticket that gives her the highest utility, or leaves if she prefers to buy nothing. If the evening screening sells out, some of its customers may switch to the afternoon screening. How should the theater price the afternoon and evening tickets to maximize the total value of all customers before knowing which customers will arrive first?

This is a unit-demand instance of a \emph{multi-unit combinatorial prophet inequality}. A seller has multiple identical copies of each item, and buyers with independent valuations drawn from known distributions arrive sequentially in an unknown fixed order. Each buyer purchases a utility-maximizing bundle of available items at the posted prices, and the seller seeks to maximize social welfare---the sum of buyers' values for their purchases---relative to the hindsight optimum (the prophet).

For a single screening (one item), this problem is well understood: a single static ticket price suffices to approach the optimal welfare as the number of seats $k$ grows~\cite{HKS07,CDL24}. With multiple screenings, however, demand for one item depends on which other items remain available. For the broad class of fractionally subadditive (XOS) valuations, \textcite{feldman2014combinatorial} showed that static item prices guarantee half the optimal welfare; their guarantee extends to any supply by treating identical copies as distinct items and pricing them identically. A natural question arises: \emph{does having many copies of each item allow static pricing to beat $1/2$ in combinatorial markets?}

One way to gain leverage from larger supply is to vary an item's price as its copies sell---for example, offering early-bird discounts on the first seats of each screening and charging more for the remaining seats. Following \textcite{ChawlaDHW25}, we study \emph{supply-based pricing}, in which the seller commits in advance to a nondecreasing price schedule for each item's copies. For XOS buyers with minimum supply $k$, \textcite{ChawlaDHW25} gave a supply-based mechanism achieving $1-(k/(k+1))^k \to 1-1/e$, and showed that fully \emph{dynamic} item pricing approaches $1$. This raises two fundamental questions: \emph{how much approximation can be achieved with any fixed number $r$ of distinct prices per item, and can unrestricted supply-based pricing approach $1$ as $k$ grows?}

\subsection{Our results}

We resolve both questions across the spectrum of price complexity, even for unit-demand buyers.

\paragraph{1. Tightness of $1/2$ for static pricing.}
In contrast to the single-item setting, where static prices approach optimal welfare as $k \to \infty$, we prove that larger supply yields no worst-case improvement for static item pricing in the combinatorial setting.

\begin{inftheorem}[Static pricing, Proposition~\ref{prop:static-half}]
For every fixed supply $k\geq 1$ and every $\epsilon>0$, there is an instance of unit-demand buyers with exactly $k$ copies of each item on which every randomized static pricing mechanism obtains at most $1/2+\epsilon$ of the expected hindsight optimum in some fixed arrival order.
\end{inftheorem}

Notably, our construction requires the number of items $m$ to be exponentially large in $k$. This dependence is unavoidable: as noted by \citet[Section~1.1]{ChawlaDHW25}, the dual prices of the ex-ante linear program with slightly reduced supplies already guarantee $1-O(\sqrt{\log m/k})$ whenever $m\geq k\geq 2$.\footnote{More generally, for $0<\epsilon<1/2$, the guarantee is $1-O(\epsilon)$ whenever $k=\Omega(\epsilon^{-2}\log(m/\epsilon))$. Scaling supplies by $1-\epsilon$ retains at least $1-\epsilon$ of the ex-ante optimum. The dual prices implement the LP's independent buyer choices until the first stockout. For each arriving buyer, the probability that earlier choices exhaust any item is at most $m\exp(-\epsilon^2 k/2)$ by a Chernoff bound, independently of her valuation.} Thus, any construction keeping the competitive ratio bounded away from $1$ as $k$ grows must have $m=\exp(\Omega(k))$.

\paragraph{2. Improved approximations for any number of distinct prices ($r\geq 2$).}
While \textcite{ChawlaDHW25} bypass the $1/2$ static barrier using up to $r=k$ price levels per item to obtain $1-(k/(k+1))^k \to 1-1/e$ for XOS buyers, using a different price for every copy is impractical.  We study how the competitive ratio depends on the number of distinct price levels $r$ per item. Surprisingly, for unit-demand buyers, adding just \emph{one} extra price per item ($r=2$) already produces a sharp jump from $1/2$ to $9/13 \approx 0.692$---surpassing $1-1/e \approx 0.632$---against the stronger ex-ante optimum. More generally, for every fixed $r \ge 2$, the limiting guarantee is $\alpha_r = \frac{1}{1+(r/(r+1))^r}$, which approaches $e/(e+1) \approx 0.731$ as $r$ grows.

\begin{inftheorem}[Few prices, Theorems~\ref{thm:two-price-guarantee} and~\ref{thm:many-price-guarantee}]
For unit-demand buyers, a randomized supply-based pricing mechanism using at most two prices per item achieves $(9/13 - o_k(1))$ of the ex-ante optimum in every fixed arrival order. More generally, for every fixed $r\geq 2$, a randomized mechanism using at most $r$ prices per item achieves $(\alpha_r - o_k(1))$ of the ex-ante optimum, where $\alpha_r = \frac{1}{1+(r/(r+1))^r}$ and $\lim_{r\to\infty}\alpha_r = e/(e+1)$. Without randomization, two prices already guarantee $2/3 - O(k^{-1/2})$ (and $r$ prices guarantee $\frac{1}{1+((r-1)/r)^{r-1}} - O(k^{-1/2} + r/k)$).
\end{inftheorem}

\paragraph{3. Supply-based pricing cannot reach $1$.}
\textcite{ChawlaDHW25} also show that \emph{dynamic} item pricing---which updates prices based on both the arriving buyer's distribution and the remaining inventory across all items---achieves $1-O(\sqrt{\log k/k})$ for XOS buyers. Is this stronger form of adaptivity necessary to approach $1$? Our third result shows that it is: even for unit-demand buyers and even when the seller may place a distinct, randomized price on every single copy, supply-based pricing remains bounded strictly away from $1$.

\begin{inftheorem}[Supply-based pricing, Proposition~\ref{prop:layered-supply}]
For every fixed $k\geq 1$ and every $\epsilon>0$, there is a unit-demand instance with $k$ copies of each item and two fixed arrival orders such that every randomized supply-based pricing mechanism achieves expected welfare at most $(2\sqrt{2}-2+\epsilon)$ times the prophet's welfare in at least one of the two orders.
\end{inftheorem}

\subsection{Technical overview}

We outline our main technical ideas for the previous three results.

\paragraph{Upper bound construction for supply-based pricing.}
Consider a main item with $k$ copies, $k/2$ low-value buyers who value only this item at $1$, and $k/2$ high-value buyers who value it at $2$, giving an optimal welfare of $3k/2$. Add a large pool of \emph{flexible} buyers, each valuing the main item at $1+U$ with $U\sim\Unif[0,\delta]$ for a tiny $\delta>0$, and compare two arrival orders. In the \emph{quiet order}, high buyers arrive first, followed by flexible buyers and then low buyers; in the \emph{congested order}, flexible buyers arrive first, followed by low buyers and then high buyers. Intuitively, preserving $k/2$ copies for high buyers in the congested order requires pricing the last $k/2$ copies above $1+\delta$ to block flexible buyers, which in turn prevents low buyers from purchasing those copies in the quiet order.

On a single item, however, the seller could still thread the needle by pricing the final $k/2$ copies right near $1+\delta$ so that enough flexible buyers purchase them in the quiet order without exhausting them in the congested order. To defeat this, we introduce a second layer of \emph{backup items}, where each flexible buyer also values a random subset of backups. Now a flexible buyer chooses the main item only if its utility exceeds the best utility among her sampled backups. In the quiet order, backups remain available and attractive, diverting flexible buyers away from the main item. In the congested order, dedicated \emph{blocker} buyers arrive first and exhaust a small fraction of the cheap backups. Because each flexible buyer samples backups independently, even a tiny drop in available cheap backups amplifies exponentially into a sharp threshold $\tau \in (1, 1+\delta)$: in the quiet order, almost no flexible buyers buy the main item at prices at least $\tau$, whereas in the congested order, flexible buyers consume every copy priced below $\tau$ before high buyers arrive. Thus no supply-based schedule can simultaneously fill the supply in the quiet order and protect high buyers in the congested order, already bounding the ratio strictly below $1$ (to $7/8$). To sharpen this constant, we extend the two-layer structure to $d$ layers---so backup items themselves face the same conflict against deeper layers---and optimize the high-to-low value ratio ($H=1+\sqrt{2}$) and order weights, yielding the $2\sqrt{2}-2$ bound for every fixed $k$.

\paragraph{Tight $1/2$ upper bound for static pricing.}
In the construction above, a supply-based seller could still reserve the last $k/2$ copies for high buyers by charging a higher price on them. Under static pricing, the first and last copies must share the same price, allowing us to tighten the upper bound to $1/2$. Recall the classic single-item, single-copy $1/2$ instance: a deterministic buyer has value $1$, followed by a buyer with high value $1+1/\eta$ with small probability $\eta$, forcing any online algorithm to lose either the base value or utility from the rare high value. We lift this tension to $k$ copies by giving a main item of base value $1$ a large pool of $N$ flexible buyers: each ordinarily values the item at $1+U$ with $U\sim\Unif[0,\delta]$, but independently draws a rare high value $1+k/\eta$ with probability $\eta/N$. The prophet always obtains $k$ by filling all copies, plus roughly another $k$ in expectation whenever a high buyer appears, for total welfare $2k$.

To force any static price to lose half of this welfare, we replace the two orders by a sequence of $L$ orders that move one additional batch of blockers ahead of the flexible buyers in each successive order. As more backup items are exhausted across orders, flexible buyers' outside options decrease; by the same independent-sampling amplification as before, ordinary demand for the main item jumps by a large multiplicative factor from one order to the next. Consequently, for any fixed static price, in every order except at most one transition order, ordinary demand is either negligible (losing the base welfare $k$) or overwhelming (exhausting all $k$ copies before rare high buyers arrive, losing the high-value $k$). Averaging over $L$ orders makes that single transition order negligible. Applying the same $d$-layer structure to the backups ensures that all but at most one layer suffer this half-welfare loss, yielding the tight $1/2$ bound for every fixed $k$.

\paragraph{Positive results for two and more price levels ($9/13$ and $\alpha_r$).}
To beat $1/2$, \textcite{ChawlaDHW25} refined the balanced-price framework of \textcite{feldman2014combinatorial} to balance seller revenue and buyer utility copy by copy. We sharpen this framework for unit-demand buyers. Consider first a deterministic two-price schedule for item $j$ that offers the first $t_j$ copies at a base price $a_j$ and the remaining copies at a higher price $b_j$. Because unit-demand buyers who prefer item $j$ at base prices $a$ never switch away when other items increase in price, we can credit \emph{both} buyer utility and seller revenue in the regime where the first $t_j$ copies do not sell out. Using Kakutani's fixed-point theorem to select base prices $a$ that equalize the welfare lower bound across all three regimes (copies left at $a_j$, copies left at $b_j$, or full sellout) yields a deterministic $2/3 - O(k^{-1/2})$ guarantee. Extending this equalization to a geometric schedule of $r$ prices gives $\bigl[1+((r-1)/r)^{r-1}\bigr]^{-1} - O(k^{-1/2}+r/k)$, approaching $e/(e+1)$ as $r$ grows.

Crucially, the deterministic two-price bound of $2/3$ is tight only when base-price demand is one-third of supply ($C_j \approx k_j/3$), leaving two-thirds of the copies unsold at the base price. To reach $9/13$, we calibrate on slightly reduced supplies $\bar k_j$ and randomize globally between two two-price schedules: with probability $10/13$, we run the calibrated schedule above; with probability $3/13$, we instead discount the first $\lfloor \bar k_j - C_j \rfloor$ copies of each item to attract extra demand while reserving the remaining copies at a slightly raised price. By setting the reserve prices via a geometrically rank-weighted utility objective and proving an exponential-moment stockout bound, we ensure that reserved copies almost never sell out---preserving nearly all base buyer utility while collecting revenue on the discounted copies. Mixing these complementary strategies shifts the worst-case demand ratio from $1/3$ to $1/2$ and improves the guarantee to $9/13 - o_k(1)$. Pairing the $r$-price geometric schedule with the same two-price discount-and-reserve strategy yields $\alpha_r = \bigl[1+(r/(r+1))^r\bigr]^{-1} - o_k(1)$ for every fixed $r\geq 2$.

\paragraph{Organization.}
Section~\ref{sec:prelim} defines the model, benchmarks, and pricing classes. Section~\ref{sec:upper-bounds} proves our impossibility results in pedagogical order: Section~\ref{sec:two-layer} introduces the backup-amplification technique in a two-layer warm-up ($7/8$), Section~\ref{sec:layered-supply} extends it to multiple layers to establish the $2\sqrt{2}-2$ supply-based upper bound, and Section~\ref{sec:static-half} combines layering with rare high values and $L$ orders to prove the tight $1/2$ bound for static pricing. Section~\ref{sec:few-prices} states our randomized guarantees ($9/13$ for two prices and $\alpha_r$ for $r$ prices) alongside our deterministic few-price mechanisms ($2/3$ for two prices and the $r$-price extension), while Appendix~\ref{app:randomized-few-prices} provides the full proofs of the randomized and $r$-price guarantees. Section~\ref{sec:related-work} discusses further related work.
\section{Preliminaries}
\label{sec:prelim}

\subsection{Combinatorial auctions and the hindsight optimum}

We consider a combinatorial auction with $m$ items and $n$ buyers.
Item $j\in[m]$ has $k_j\in\Z_{>0}$ identical copies available.
We write $K=(k_1,\ldots,k_m)$ for the supply vector and $k=\min_{j\in[m]}k_j$ for the minimum multiplicity.
Each buyer can receive at most one copy of each item, so we identify her allocation with a subset of $[m]$.
Buyer $i$ has a valuation $v_i:2^{[m]}\to\R_{\geq0}$ drawn from a known independent distribution $\D_i$.
Let $\D=\D_1\times\cdots\times\D_n$ denote the joint distribution.
An instance is specified by $(\D,K)$.
We assume that valuations are normalized, $v_i(\emptyset)=0$, and monotone, $v_i(S)\leq v_i(T)$ whenever $S\subseteq T$.

Our goal is to design a mechanism that maximizes the social welfare. For a realized valuation profile $\vals=(v_1,\ldots,v_n)$, the hindsight optimal social welfare is
\begin{equation}
\begin{aligned}
\hopt(\vals,K)\coloneqq
\max_{S_1,\ldots,S_n\subseteq[m]}\quad
    &\sum_{i\in[n]}v_i(S_i)\\
\text{subject to}\quad
    &\sum_{i\in[n]}\one\{j\in S_i\}\leq k_j
        &&\forall j\in[m].
\end{aligned}
\tag{HOpt}\label{eq:hindsight-opt}
\end{equation}
The prophet's reward is its expectation:
\[
    \prophet(\D,K)\coloneqq\E_{\vals\sim\D}\bigl[\hopt(\vals,K)\bigr].
\]

We also consider the stronger ex-ante relaxation, in which the supply constraints hold in expectation.
For each buyer $i$ and valuation $v\in\operatorname{supp}(\D_i)$, let $x_{i,v,S}$ be the probability of assigning her set $S$ conditional on her valuation being $v$.
These conditional probabilities are measurable functions of $v$.
The ex-ante optimum is
\begin{equation}
\begin{aligned}
\eaopt(\D,K)\coloneqq\max_{x\geq0}\quad
    &\sum_{i\in[n]}\sum_{S\subseteq[m]}
        \E_{v_i\sim\D_i}\bigl[v_i(S)x_{i,v_i,S}\bigr]\\
\text{subject to}\quad
    &\sum_{i\in[n]}\sum_{S\subseteq[m]:\,j\in S}
        \E_{v_i\sim\D_i}\bigl[x_{i,v_i,S}\bigr]\leq k_j
        &&\forall j\in[m],\\
    &\sum_{S\subseteq[m]}x_{i,v,S}\leq1
        &&\forall i,v.
\end{aligned}
\tag{EA-Opt}\label{eq:ex-ante-opt}
\end{equation}
Any hindsight allocation induces a feasible solution to \eqref{eq:ex-ante-opt} by conditioning on each buyer's valuation.
Consequently, $\eaopt(\D,K)\geq\prophet(\D,K)$.
Our positive guarantees use the stronger ex-ante benchmark, whereas our impossibility results use the weaker prophet benchmark.

\subsection{Buyer valuations}

We consider the following valuation classes.
\begin{itemize}[leftmargin=*,itemsep=4pt]
    \item \textbf{Unit demand.}
    A valuation is unit demand if there are nonnegative item values $v_j$ such that
    $v(S)=\max_{j\in S}v_j$, with $v(\emptyset)=0$.

    \item \textbf{Fractionally subadditive (XOS).}
    A valuation is XOS if there is a collection $\mathcal A_v\subseteq\R_{\geq0}^m$ of additive valuations such that
    \[
        v(S)=\max_{a\in\mathcal A_v}\sum_{j\in S}a_j
        \qquad\text{for every }S\subseteq[m].
    \]
    For any set $S$, an additive valuation $a^{v,S}$ attaining this maximum is called a \emph{supporting additive valuation} for $S$.
    It satisfies $\sum_{j\in S}a^{v,S}_j=v(S)$ and $\sum_{j\in T}a^{v,S}_j\leq v(T)$ for every $T\subseteq[m]$.
\end{itemize}
Every unit-demand valuation is XOS.

\subsection{Pricing mechanisms}

In this work, we focus on \emph{pricing} mechanisms.
Buyers arrive sequentially. Whenever a buyer arrives, the mechanism posts a price for each item that still has remaining copies, and the buyer purchases a utility-maximizing bundle of available items at the posted prices (including $\emptyset$).
\footnote{Our upper bounds in Section~\ref{sec:upper-bounds} hold under every tie-breaking rule, whereas our positive guarantees in Section~\ref{sec:few-prices} assume generic distributions where ties occur with probability zero (see Section~\ref{sec:few-prices} for details).}

We study two classes of pricing mechanisms, following the terminology of \textcite{ChawlaDHW25}.

\paragraph{Static item pricing.}
The seller chooses a nonnegative price $p_j$ for each item $j$ before buyers arrive.
Every available copy of item $j$ is offered at this price throughout the process.

\paragraph{Supply-based item pricing.}
The seller chooses a nonnegative price schedule $p_{j,1},\ldots,p_{j,k_j}$ for each item $j$ before buyers arrive.
After $c$ copies of item $j$ have been sold, its next copy is offered at price $p_{j,c+1}$.
Thus, an item's price depends only on the number of its own copies already sold.
We consider \emph{nondecreasing} schedules, satisfying $p_{j,1}\leq\cdots\leq p_{j,k_j}$.
A schedule uses \emph{at most $r$ prices per item} if $|\{p_{j,1},\ldots,p_{j,k_j}\}|\leq r$ for every $j\in[m]$; in particular, $r=1$ recovers static item pricing.

\paragraph{Randomization and withholding.}
A \emph{randomized} static or supply-based pricing mechanism draws its price vector or price schedules from a probability distribution before any buyer arrives, independently of the realized valuations and the arrival order.
Our impossibility results in Section~\ref{sec:upper-bounds} also allow the seller to withhold arbitrary copies of each item before buyers arrive (equivalently, to set prices in $\R_{\geq 0}\cup\{\infty\}$).

\subsection{Performance metric}

The seller knows $(\D,K)$ but not the arrival order $\pi$.
An adversary fixes $\pi$ knowing the instance and the mechanism, independently of the realized valuations and the mechanism's private randomness.

For a fixed arrival order $\pi$, let
\[
    \welfare(M,\D,K;\pi)
    \coloneqq\E\left[\sum_{i\in[n]}v_i(S_i)\right],
\]
where the expectation is over valuations and the mechanism's randomness.
A prior-dependent mechanism $M$ is $\alpha(k)$-competitive against the ex-ante benchmark if, for every instance $(\D,K)$ of minimum multiplicity at least $k$ and every fixed order $\pi$,
\[
    \welfare(M,\D,K;\pi)
    \geq\alpha(k)\,\eaopt(\D,K).
\]
Equivalently, its competitive ratio is
\[
    \compratio(M,k)
    \coloneqq
    \inf_{\substack{\D,K:\,\min_j k_j\geq k}}
    \inf_{\pi}
    \frac{\welfare(M,\D,K;\pi)}{\eaopt(\D,K)}.
\]
The infimum is taken over the valuation class specified in the theorem.

When we write an error term $o_k(1)$, its convergence depends only on $k$ and any explicitly fixed mechanism parameter (such as the number of price levels $r$), and not on the number of items $m$, the number of buyers $n$, the value range, or the supports of the priors.

\section{Upper Bounds}
\label{sec:upper-bounds}

In this section, we show that for every fixed supply $k\geq1$, static pricing cannot guarantee a competitive ratio greater than $1/2$, and unrestricted supply-based pricing cannot guarantee a ratio greater than $2\sqrt2-2$.
All of our constructions use only unit-demand buyers and allow the seller to randomize prices and withhold copies before buyers arrive.
By a general reduction proved in Appendix~\ref{app:initial-buyers} (Lemma~\ref{lem:positive-prices}), we may assume without loss of generality in each construction that every offered copy has a strictly positive price---so every unit-demand buyer purchases at most one item under any tie-breaking rule---at the cost of an arbitrarily small additive welfare loss $\gamma>0$.

We develop the proofs in three steps.
Section~\ref{sec:two-layer} gives a warm-up two-layer example that uses random backup items to create conflicting demands for one main item across two arrival orders, yielding a $7/8$ bound.
Section~\ref{sec:layered-supply} extends this conflict to multiple layers so that backup items face the same tradeoff and optimizes the value ratio and order weights, bringing the bound down to $2\sqrt2-2$ for every fixed $k$.
Finally, Section~\ref{sec:static-half} combines layering with rare high values and $L$ congestion orders to obtain the tight $1/2$ bound for static pricing at every fixed $k$.

\subsection{Warm-up: a two-layer construction}
\label{sec:two-layer}

We first describe our intuition for constructing the instance. Our example consists of one main item and many backup items, each with $k$ copies. For the main item, $k/2$ buyers have value $2$ while $k/2$ buyers have value $1$. The prophet obtains welfare $3k/2$ by serving all these buyers. There are also many flexible buyers who value the main item slightly above $1$. Each flexible buyer has a positive value for a random set of backup items. Additional buyers value only the backup items; their purchases alter the utility that flexible buyers obtain from the backup items.

We compare the performance of any supply-based pricing mechanism in two fixed arrival orders. In the \emph{quiet order}, the high-value buyers arrive first and the backups remain attractive to the flexible buyers.
In the \emph{congested order}, buyers for the backup items arrive first, reducing the flexible buyers' utility from backup items; the high-value buyers arrive last. The construction amplifies this change in outside options into a large change in demand for the main item. Too few low-priced copies leave supply unsold in the quiet order, while too many let flexible buyers consume copies needed by the high-value buyers in the congested order. No single supply-based schedule can perform well in both orders.

We now state the main results of this subsection.

\begin{theorem}
\label{thm:two-layer}
For every integer $k\geq2$, there is an instance with unit-demand buyers, exactly $k$ copies of every item, and two fixed arrival orders $\pi_Q,\pi_B$, such that every supply-based pricing mechanism $M$ satisfies
\[
    \frac{\welfare(M,\D,K;\pi_Q)+\welfare(M,\D,K;\pi_B)}{2}
    \leq\left(\frac78+O\!\left(\frac1k\right)\right)
        \prophet(\D,K).
\]
\end{theorem}

\begin{proposition}
\label{prop:two-layer}
For every $\epsilon>0$ and integer $k_0\geq1$, there is an instance of unit-demand buyers with exactly $k\geq k_0$ copies of every item such that, for every supply-based pricing mechanism $M$, some fixed arrival order $\pi$ satisfies
\[
    \welfare(M,\D,K;\pi)
    \leq\left(\frac78+\epsilon\right)\prophet(\D,K).
\]
Thus, even with arbitrarily large minimum supply, supply-based pricing cannot guarantee a competitive ratio strictly larger than $7/8$.
\end{proposition}

\begin{proof}[Proof of Theorem~\ref{thm:two-layer}]
By Lemma~\ref{lem:positive-prices} (with $\gamma=1/4$), it suffices to construct a unit-demand instance with $k$ copies of every item, prophet welfare at least $2k$, and two fixed arrival orders whose average welfare is at most $7k/4+15/4$ for every deterministic supply-based mechanism with strictly positive offered prices.
The bound of $7k/4+4$ then follows for every deterministic mechanism on the augmented instance from Lemma~\ref{lem:positive-prices}, and extends to randomized mechanisms by averaging over their random choices.

\paragraph{The instance.}
Fix $k\geq2$.
There is one main item and $m$ backup items.
Every item has $k$ copies.
Let $a=1/(2m)$ and $\delta=a/k^3$.
We also use integer parameters $T,N,m$ that depend only on $k$ and are specified in Appendix~\ref{app:two-layer-amplification}.
There are four groups of buyers; values for items not specified are zero.
\begin{itemize}[leftmargin=*,itemsep=3pt]
    \item \emph{Main-item buyers:} $k/2$ high buyers value only the main item at $2$, and $k/2$ low buyers value only the main item at $1$.\footnote{If $k$ is odd, we instead have $(k-1)/2$ high and low buyers and one more buyer whose value is high or low with probability $1/2$ each. The remaining analysis works the same.}
    \item \emph{Flexible buyers:} each of $N$ buyers independently draws $U\sim\Unif[0,\delta]$ and samples $T$ backup items uniformly with replacement, independently of $U$.
    She values the main item at $1+U$ and each sampled backup at $2a$.
    Repeated samples do not increase an item's value.
    \item \emph{Backup low buyers:} for each backup item, $k$ buyers value only that item at $a$.
    \item \emph{Blockers:} for each backup item, $k$ additional buyers independently value only that item at $4a$ with probability $1/k$, and otherwise have the zero valuation.
\end{itemize}

\paragraph{Arrival orders and the benchmark.}
Fix an internal order for each group.
The two orders are
\[
\begin{aligned}
    \pi_Q:\quad&\text{highs}\ \to\ \text{flexible buyers}\ \to\ \text{lows}
                  \ \to\ \text{backup lows}\ \to\ \text{blockers},\\
    \pi_B:\quad&\text{backup lows}\ \to\ \text{blockers}
                  \ \to\ \text{flexible buyers}\ \to\ \text{lows}\ \to\ \text{highs}.
\end{aligned}
\]

The prophet can ignore the flexible buyers and blockers and serve all main-item buyers and backup low buyers at their respective items.
Therefore,
\begin{equation}
    \prophet(\D,K)\geq
    2\cdot\frac k2+\frac k2+mka=2k.
    \label{eq:two-layer-benchmark}
\end{equation}

\paragraph{The pricing tradeoff.}
Consider any deterministic supply-based pricing mechanism whose offered prices are all strictly positive, so every buyer purchases at most one item.
The initial schedules are the same in both orders.
For each available item, call its price at the start of the order its \emph{initial price}.
Call a backup item \emph{cheap} if it is initially available at a price at most $a$.

We use the following claim to compare the behavior of the flexible buyers in the two orders.
\begin{claim}
\label{clm:two-layer-amplification}
The parameters can be chosen so that $N/m\leq1$ and the following holds for every deterministic supply-based mechanism with strictly positive offered prices.
If at least $m/k$ backups are cheap, there is a threshold $\tau\in(1,1+\delta)$ such that:
\begin{enumerate}[label=(\roman*),leftmargin=*,itemsep=3pt]
    \item In the quiet order, the expected number of flexible purchases of the main item at prices at least $\tau$ is at most $1$.
    \item In the congested order, with probability at least $1-k^{-2}$, every copy of the main item with price below $\tau$ is consumed before the high buyers arrive.
\end{enumerate}
\end{claim}

\paragraph{Intuition for Claim~\ref{clm:two-layer-amplification}.}
A flexible buyer compares her utility from the main item with the best utility among her $T$ sampled backups.
In the quiet order, earlier flexible buyers can have bought copies of at most $N$ distinct backups. Every other initially available backup is still offered at its initial price.
In the congested order, each initially cheap backup has a small independent probability that all its blockers are active.
When this happens, the backup either sells out or becomes too expensive for a flexible buyer.
In the congested order, an additional fraction of the backup items become expensive, and since a buyer samples independently, the probability that all her backups are expensive is the $T$th power of the corresponding fraction.
Taking $T$ large enough amplifies this small change in the fraction of expensive backup items into a large change in demand for the main item.
We choose $\tau$ so that purchases at prices at least $\tau$ are rare in the quiet order, while the same threshold causes flexible buyers in the congested order to consume every copy priced below it.
Appendix~\ref{app:two-layer-amplification} gives the parameter choices and probability bounds.

\paragraph{Welfare assuming Claim~\ref{clm:two-layer-amplification}.}
We first derive the welfare bound from the claim.
First suppose that at least $m/k$ backups are cheap.
Let $s$ be the number of copies of the main item offered for sale at prices below $\tau$.
Let $W^Q$ and $W^B$ denote the realized welfare from purchases of the main item in the respective orders.
In the quiet order, if all $k/2$ high buyers purchase the main item, they buy $k/2$ copies and contribute welfare $k$.
At most $(s-k/2)_+$ further purchases can occur below $\tau$.
Since $\tau>1$, no low buyer can purchase at a price of $\tau$ or above, and by part~(i) of Claim~\ref{clm:two-layer-amplification} at most $1$ flexible buyer does so in expectation.
If some high buyer does not purchase, the current price must be at least $2$, so the subsequent flexible and low buyers cannot purchase either.
Thus
\begin{equation}
    \E[W^Q]\leq
    \frac k2+\max\{s,k/2\}+1+k\delta.
    \label{eq:two-layer-quiet}
\end{equation}
Here $k\delta$ accounts for all additional welfare of flexible buyers above value $1$.

In the congested order, on the event in part~(ii) of the claim, at most $k-s$ copies remain for the high buyers.
Every other buyer's value for the main item is at most $1+\delta$.
Counting a base value of $1$ for each of the $k$ copies and an additional $1$ for each high-buyer purchase yields
\begin{equation}
    \E[W^B]\leq
    k+\min\{k/2,k-s\}+k\delta+\frac1{2k}.
    \label{eq:two-layer-busy}
\end{equation}
The last term bounds the additional high-buyer welfare on the event where part~(ii) of Claim~\ref{clm:two-layer-amplification} does not hold.
We have
\[
    \max\{s,k/2\}+\min\{k/2,k-s\}=k.
\]

Averaging \eqref{eq:two-layer-quiet} and \eqref{eq:two-layer-busy} gives
\begin{equation}
    \frac{\E[W^Q]+\E[W^B]}{2}
    \leq\frac{5k}{4}+\frac12+k\delta+\frac1{4k}.
    \label{eq:two-layer-main-average}
\end{equation}
The leading main-item contribution is $5k/4$, compared with its benchmark contribution of $3k/2$.
Adding at most $mka=k/2$ from backup low buyers, $2$ in expectation from blockers, and $N/m\leq1$ from flexible backup purchases, the total two-order average is at most
\[
    \frac{7k}{4}+\frac72+k\delta+\frac1{4k}
    \leq\frac{7k}{4}+\frac{15}{4}.
\]
The last inequality uses $\delta=1/(2mk^3)$, $m\geq1$, and $k\geq2$.

It remains to consider mechanisms for which fewer than $m/k$ backups are cheap.
Only initially cheap backups can be purchased by their low buyers under a supply-based schedule.
Their total welfare is therefore at most $(m/k)ka=1/2$.
Even granting the main item its full possible welfare $3k/2+k\delta$, expected welfare in either order is at most
\[
    \frac{3k}{2}+\frac72+k\delta
    \leq\frac{7k}{4}+\frac{15}{4}.
\]
Applying Lemma~\ref{lem:positive-prices} with $\gamma=1/4$ gives an augmented unit-demand instance with two fixed arrival orders on which every deterministic supply-based mechanism achieves two-order average welfare at most $7k/4+4$, while $\prophet(\D,K)\geq2k$ by \eqref{eq:two-layer-benchmark}.
Averaging over a randomized mechanism's choices preserves this bound, yielding
\begin{equation}
    \frac{\welfare(M,\D,K;\pi_Q)+\welfare(M,\D,K;\pi_B)}
         {2\prophet(\D,K)}
    \leq\frac{7k/4+4}{2k}
    =\frac78+\frac2k.
    \label{eq:two-layer-final-ratio}
\end{equation}
This proves Theorem~\ref{thm:two-layer} assuming Claim~\ref{clm:two-layer-amplification}. Choosing $k\geq\max\{k_0,\lceil2/\epsilon\rceil\}$ and taking the worse of the two orders then gives Proposition~\ref{prop:two-layer}.
\end{proof}

\subsection{A $2\sqrt2-2$ bound for supply-based pricing}
\label{sec:layered-supply}

The two-layer example creates a conflict between selling enough copies in the quiet order and preserving copies for high buyers in the congested order.
Its welfare calculation uses the two orders only for the main item and gives the backup low buyers their full benchmark contribution.
We now make the backup items face the same conflict: each receives its own high, low, and flexible buyers, with items in later layers serving as backups.
Every layer then contributes equally to the benchmark, and all but at most one layer satisfy the same welfare bound.

We also adjust the probabilities of the two orders, which were both $1/2$ in the two-layer example.
For an item with low value $a$ and high value $Ha$, selling one more copy to a low or flexible buyer contributes a base value of $a$ in the quiet order. In the congested order, a copy bought by a high buyer contributes an additional $(H-1)a$ above its base value.
We give these contributions equal weight by choosing probabilities $w=(H-1)/H$ and $1-w=1/H$ for the quiet and congested orders.
As we show below, the resulting weighted welfare is at most $ka(1+w(H-1)/2)$ up to the error in the construction. Dividing by the benchmark $(H+1)ka/2$ gives the leading term
\[
    \frac{2+w(H-1)}{H+1}=1-\frac{H-1}{H(H+1)}.
\]
Choosing $H=1+\sqrt2$ minimizes the leading term at $2\sqrt2-2$ and gives probabilities $2-\sqrt2$ and $\sqrt2-1$ for the two orders.

To obtain this bound while keeping $k$ fixed, we choose the number of layers and the amplification parameters using a separate error parameter $\eta$.

\begin{samepage}
\begin{theorem}
\label{thm:layered-supply}
For every integer $k\geq1$ and $\eta\in(0,1/8)$, there is an instance with unit-demand buyers, exactly $k$ copies of every item, and two fixed arrival orders $\pi_Q,\pi_B$, such that every supply-based pricing mechanism $M$ satisfies
\[
    w\,\welfare(M,\D,K;\pi_Q)
    +(1-w)\,\welfare(M,\D,K;\pi_B)
    \leq\left(2\sqrt2-2+5\eta\right)
        \prophet(\D,K),
\]
where $w=2-\sqrt2$.
\end{theorem}
\end{samepage}

\begin{samepage}
\begin{proposition}
\label{prop:layered-supply}
For every integer $k\geq1$ and $\epsilon>0$, there is an instance of unit-demand buyers with exactly $k$ copies of every item such that, for every supply-based pricing mechanism $M$, some fixed arrival order $\pi$ satisfies
\[
    \welfare(M,\D,K;\pi)
    \leq\left(2\sqrt2-2+\epsilon\right)\prophet(\D,K).
\]
Thus, for every fixed supply $k$, supply-based pricing cannot guarantee a competitive ratio strictly larger than $2\sqrt2-2$.
\end{proposition}
\end{samepage}

\begin{proof}[Proof of Theorem~\ref{thm:layered-supply}]
Fix $k\geq1$ and $\eta\in(0,1/8)$.
By Lemma~\ref{lem:positive-prices} (with $\gamma=\eta k/2$), it suffices to construct a unit-demand instance with $k$ copies of every item and two fixed arrival orders $\pi_Q,\pi_B$ for which every deterministic supply-based pricing mechanism with strictly positive offered prices satisfies the weighted welfare bound below up to an additive margin of $\eta k/2$.
The bound then follows for every deterministic mechanism on the augmented instance from Lemma~\ref{lem:positive-prices}, and extends to randomized mechanisms by averaging over their random choices.

\paragraph{The instance.}
Let
\[
    d=\lceil1/\eta\rceil,\qquad H=1+\sqrt2.
\]
We also use integer parameters $m_1,R,N,T$ that depend only on $k$ and $\eta$ and are chosen in Appendix~\ref{app:layered-supply}.
There are $d$ layers.
Layer $\ell$ contains $m_\ell=m_1R^{\ell-1}$ items, each with $k$ copies and base value $a_\ell=1/m_\ell$.
In particular, $m_\ell a_\ell=1$ for every layer.
Set $\delta=\eta a_d/4$.
For each item $j$ in layer $\ell$, introduce the following buyers, with all unspecified values equal to zero.
\begin{itemize}[leftmargin=*,itemsep=3pt]
    \item \emph{High and low buyers:} $k/2$ high buyers value only $j$ at $Ha_\ell$, and $k/2$ low buyers value only $j$ at $a_\ell$. (For odd $k$, use $(k-1)/2$ buyers of each kind and one more buyer, last in the high group in both orders, whose value is high or low with equal probability.\footnote{The number $X$ of high buyers then takes the two integers on either side of $k/2$ with equal probability. For any integer cutoff $s$, averaging $\max\{s,X\}$ and $\min\{X,k-s\}$ gives $\max\{s,k/2\}$ and $\min\{k/2,k-s\}$, respectively.} We give the calculation for even $k$.)
    \item \emph{Flexible buyers:} if $\ell<d$, each of $N$ buyers independently draws $U\sim\Unif[0,\delta]$ and values her main item $j$ at $a_\ell+U$.
    Independently, she makes $T$ backup draws.
    Each draw chooses a layer $t$ uniformly from $\{\ell+1,\ldots,d\}$ and then an item uniformly from that layer.
    She values every sampled layer-$t$ item at $2a_t$; repeated samples do not increase its value.
    \item \emph{Blockers:} each of $k$ buyers independently values only $j$ at $4a_\ell$ with probability $\eta$, and otherwise has the zero valuation.
\end{itemize}

\paragraph{Arrival orders and the benchmark.}
Write $\mathsf H_\ell,\mathsf F_\ell,\mathsf L_\ell,\mathsf Z_\ell$ for the high, flexible, low, and blocker groups of layer $\ell$, and fix an internal order for each group.
We use
\[
\begin{array}{lll}
    \pi_Q:&\text{layers }1,2,\ldots,d:&
        \mathsf H_\ell\ \to\ \mathsf F_\ell\ \to\ \mathsf L_\ell\ \to\ \mathsf Z_\ell,\\[2pt]
    \pi_B:&\text{layers }d,d-1,\ldots,1:&
        \mathsf F_\ell\ \to\ \mathsf L_\ell\ \to\ \mathsf H_\ell\ \to\ \mathsf Z_\ell.
\end{array}
\]
In the quiet order, a flexible buyer arrives before the buyers assigned to all her backup layers.
In the congested order, those layers, including their blockers, have already been processed.
Choose the quiet order with probability $w=(H-1)/H=2-\sqrt2$ and the congested order with probability $1-w$.

The prophet can serve all high and low buyers at their respective items. Each layer contributes $(H+1)k/2$ since $m_\ell a_\ell=1$, so
\begin{equation}
    \prophet(\D,K)\geq \frac{d(H+1)k}{2}.
    \label{eq:layered-benchmark}
\end{equation}

\paragraph{The pricing tradeoff.}
Consider any deterministic supply-based pricing mechanism whose offered prices are all strictly positive, so every buyer purchases at most one item.
Call an item in layer $\ell$ \emph{cheap} if it is initially available at a price at most $a_\ell+\delta$.
Let $c_\ell$ be the fraction of cheap items in that layer.
The following claim extends Claim~\ref{clm:two-layer-amplification} to buyers whose backups may belong to several later layers.

\begin{claim}
\label{clm:layered-amplification}
The parameters can be chosen so that the following holds for every deterministic supply-based mechanism with strictly positive offered prices.
If some layer $t>\ell$ has $c_t\geq1/4$, then, for every item $j$ in layer $\ell$, there is a threshold $\tau_j\in(a_\ell,a_\ell+\delta)$ such that:
\begin{enumerate}[label=(\roman*),leftmargin=*,itemsep=3pt]
    \item In the quiet order, the expected number of purchases of $j$ at prices at least $\tau_j$ by its own flexible buyers is at most $k\eta$.
    \item In the congested order, with probability at least $1-\eta$, every copy of $j$ offered at a price below $\tau_j$ is sold before its own high buyers arrive.
\end{enumerate}
\end{claim}

The proof of Claim~\ref{clm:layered-amplification} uses the same amplification argument as Claim~\ref{clm:two-layer-amplification} and is deferred to Appendix~\ref{app:layered-supply}.

\paragraph{Welfare assuming Claim~\ref{clm:layered-amplification}.}
Fix an item $j$ in layer $\ell$ to which the claim applies, and let $s$ be the number of copies offered below $\tau_j$. As in Section~\ref{sec:two-layer}, write $W_j^Q,W_j^B$ for the welfare from $j$'s own high, low, and flexible buyers purchasing $j$ in the two orders.
Count $a_\ell$ for each sale, an additional $(H-1)a_\ell$ for each high buyer, and at most $k\delta$ in total for the flexible buyers' values above $a_\ell$.

In the quiet order, set aside the flexible purchases at prices at least $\tau_j$. If the price reaches at least $\tau_j$ before all high buyers finish arriving, then no subsequent low buyer can purchase (since $\tau_j>a_\ell$), leaving at most $k/2$ counted sales. Otherwise every counted sale occurs at a price below $\tau_j$, giving at most $s$ counted sales. Combining this with part~(i) yields at most $\max\{s,k/2\}+k\eta$ sales in expectation.
In the congested order, count a base value of $a_\ell$ for all $k$ copies. By part~(ii), at most $\min\{k/2,k-s\}$ high buyers can purchase except on an event of probability at most $\eta$. Thus
\[
\begin{aligned}
 \E[W_j^Q]&\leq a_\ell\bigl(\max\{s,k/2\}+(H-1)k/2+k\eta\bigr)+k\delta,\\
 \E[W_j^B]&\leq a_\ell\bigl(k+(H-1)\min\{k/2,k-s\}+(H-1)k\eta\bigr)+k\delta.
\end{aligned}
\]
Since $(1-w)(H-1)=w$, the maximum and minimum enter the weighted average with the same coefficient. Their sum is $k$, just as in Section~\ref{sec:two-layer}. Using $1+w(H-1)/2=\sqrt2$ gives
\begin{equation}
    w\E[W_j^Q]+(1-w)\E[W_j^B]\leq a_\ell k(\sqrt2+2w\eta)+k\delta .
    \label{eq:layered-curve}
\end{equation}

Set aside the $k\delta$ per item for the final sum. Let $\ell_*$ be the last layer with at least a quarter of its items cheap, if one exists. Each earlier layer contributes at most $k(\sqrt2+2w\eta)$ by \eqref{eq:layered-curve} and $m_\ell a_\ell=1$.
In each later layer, only high buyers can purchase their own noncheap items (since every low or flexible buyer values $j$ at most $a_\ell+\delta$), contributing at most $Ha_\ell k/2$ per such item. Even allowing $(H+1)ka_\ell/2$ per cheap item, the layer contributes at most $\tfrac34\cdot\tfrac{Hk}{2}+\tfrac14\cdot\tfrac{(H+1)k}{2}=Hk/2+k/8\leq\sqrt2\,k$. We count the full $(H+1)k/2$ for layer $\ell_*$. If no layer has a quarter of its items cheap, the later-layer estimate applies to every layer.

Blockers contribute at most $4\eta kd$ in expectation, and flexible buyers purchasing backups contribute at most $2N(d-1)/R\leq\eta kd$. By \eqref{eq:layered-size-bounds}, the $k\delta$ per item contributes at most $k\delta\sum_{\ell=1}^d m_\ell\leq\eta k/2$. Summing all contributions gives a weighted average welfare of at most
\[
    (d-1)\sqrt2\,k+\frac{(H+1)k}{2}+7\eta kd+\frac{\eta k}{2}.
\]
Applying Lemma~\ref{lem:positive-prices} with $\gamma=\eta k/2$ yields an augmented unit-demand instance and two fixed arrival orders on which every deterministic supply-based mechanism satisfies
\begin{equation}
    w\,\welfare(M,\D,K;\pi_Q)+(1-w)\,\welfare(M,\D,K;\pi_B)
    \leq (d-1)\sqrt2\,k+\frac{(H+1)k}{2}+7\eta kd+\eta k .
    \label{eq:layered-total}
\end{equation}
Here $2\sqrt2/(H+1)=2\sqrt2-2$ and $(H+1)k/2=\sqrt2\,k+wk/2$. Dividing by the benchmark $d(H+1)k/2$ from \eqref{eq:layered-benchmark} and using $1/d\leq\eta$ bounds the ratio by $2\sqrt2-2+(w+16)\eta/(H+1)<2\sqrt2-2+5\eta$.
Averaging over a randomized mechanism's choices proves Theorem~\ref{thm:layered-supply}. Taking $\eta<\epsilon/5$ and choosing the worse of the two orders gives Proposition~\ref{prop:layered-supply}.
\end{proof}

\subsection{A $1/2$ bound for static pricing}
\label{sec:static-half}

For static pricing, we keep the layers of Section~\ref{sec:layered-supply} and move one more batch of blockers ahead of the flexible buyers in each successive order. With fewer backups available, more flexible buyers may prefer their main item at the same price. We give each flexible buyer a small chance of a high value for her main item and approach the $1/2$ bound by increasing the numbers of orders and layers while keeping $k$ fixed.

\begin{theorem}
\label{thm:static-half}
For every integer $k\geq1$ and $\eta\in(0,1/8)$, there is an instance of unit-demand buyers with exactly $k$ copies of every item, and $L=\lceil1/\eta\rceil$ fixed arrival orders $\pi_0,\ldots,\pi_{L-1}$, such that every static pricing mechanism $M$ satisfies
\[
    \frac1L\sum_{b=0}^{L-1}\welfare(M,\D,K;\pi_b)
    \leq\left(\frac12+5\eta\right)\prophet(\D,K).
\]
\end{theorem}

\begin{proposition}
\label{prop:static-half}
For every integer $k\geq1$ and $\epsilon>0$, there is an instance of unit-demand buyers with exactly $k$ copies of every item such that, for every static pricing mechanism $M$, some fixed arrival order $\pi$ satisfies
\[
    \welfare(M,\D,K;\pi)
    \leq\left(\frac12+\epsilon\right)\prophet(\D,K).
\]
Thus static pricing cannot guarantee a competitive ratio strictly larger than $1/2$ for any fixed supply $k$.
\end{proposition}

\begin{proof}[Proof of Theorem~\ref{thm:static-half}]
Fix $k\geq1$ and $\eta\in(0,1/8)$.
By Lemma~\ref{lem:positive-prices} (with $\gamma=k\delta G$, for $\delta$ and $G$ defined below), it suffices to construct a unit-demand instance with $k$ copies of every item and $L$ fixed arrival orders for which every deterministic static pricing mechanism (allowing withholding) with strictly positive offered prices satisfies the average welfare bound below up to an additive margin of $k\delta G$.
The bound then follows for every deterministic mechanism on the augmented instance from Lemma~\ref{lem:positive-prices}, and extends to randomized mechanisms by averaging over their random choices.

\paragraph{The instance.}
Let $d=L=\lceil1/\eta\rceil$ be the number of layers and the number of orders. We also use integer parameters $\lambda,N$ that depend only on $k$ and $\eta$ and are specified in Appendix~\ref{app:static-half}. Layer $\ell$ has $m_\ell=N^{2\ell}$ items of base value $a_\ell=1/m_\ell$, each with $k$ copies. Write $G=\sum_{\ell=1}^d m_\ell$ and $\delta=\eta a_d$. We introduce two groups of buyers, with all unspecified values equal to zero.
\begin{itemize}[leftmargin=*,itemsep=3pt]
    \item \emph{Flexible buyers:} each layer-$\ell$ item $j$ has $N$ designated buyers. Each independently draws a high value $V_\ell=(1+k/\eta)a_\ell$ for $j$ with probability $\eta/N$, and otherwise an ordinary value $a_\ell+U$, where $U\sim\Unif[0,\delta]$. Independently, she includes each item of every later layer $s>\ell$ as a backup with probability $\lambda/m_s$, valuing it at $2a_s$.
    \item \emph{Blockers:} each layer-$\ell$ item has $L-1$ batches of $k$ buyers. Each independently values only that item at $3a_\ell$ with probability $\alpha=\eta/(8k(L-1))$, and otherwise has the zero valuation.
\end{itemize}

\paragraph{Arrival orders and the benchmark.}
Order $\pi_b$ consists of the first $b$ blocker batches of each item, followed by the flexible buyers in increasing layer order, and then the remaining blockers. Fix the internal order within these groups and choose $b$ uniformly from $0,\ldots,L-1$.

The prophet can fill each item with its designated buyers, including a high buyer whenever one appears. This happens with probability $1-(1-\eta/N)^N\geq1-e^{-\eta}\geq\eta-\eta^2/2$. Since $m_\ell a_\ell=1$, we obtain
\begin{equation}
    \prophet(\D,K)\geq\sum_{\ell=1}^d m_\ell
    \bigl[ka_\ell+(\eta-\eta^2/2)(V_\ell-a_\ell)\bigr]
    =2dk(1-\eta/4).
    \label{eq:static-benchmark}
\end{equation}

\paragraph{Demand across the orders.}
Fix a deterministic static pricing mechanism whose offered prices are all strictly positive (so every buyer purchases at most one item), and use the same blocker values in every order. Call an initially offered item in layer $\ell$ \emph{cheap} if its price is at most $a_\ell+\delta$. Let $\ell_*$ be the last layer with at least an $\eta$ fraction of cheap items, if one exists.

For an offered item $j$ with price $p_j$, imagine offering a copy just after the early blockers of $\pi_b$, even if they bought its last copy. Let $D_{j,b}$ be $N$ times the probability that a designated buyer with an ordinary value strictly prefers $j$ to buying nothing and to all her backups still offered then.

\begin{claim}
\label{clm:static-demand}
For every blocker realization, the expected number of ordinary buyers purchasing their own item $j$ in $\pi_b$ is at most $2D_{j,b}$. If $\ell_*$ exists, then with probability at least $1-\eta$ over the blockers, every offered item $j$ in a layer $\ell<\ell_*$ has at most one order with $\eta/2<D_{j,b}<4k/\eta$.
\end{claim}

At a price $p_j\leq a_\ell$, buyers who draw no backups already give large demand. At a higher price, any cheap backup in layer $\ell_*$ offers more utility than $j$. Each extra batch of early blockers removes some of those backups. A buyer includes each backup independently, so the probability of avoiding every attractive backup grows by a large factor from one order to the next. This is the amplification from Section~\ref{sec:layered-supply}, now used across $L$ orders. The formal proof of the claim is deferred to Appendix~\ref{app:static-half}.

\begin{claim}
\label{clm:static-competition}
Conditional on any blocker realization, if $D_{j,b}>0$, the expected number of high-value buyers purchasing their own item $j$ in $\pi_b$ is at most
\[
    \frac{k\eta}{(1-\eta/N)D_{j,b}} .
\]
\end{claim}

An ordinary buyer who preferred $j$ just after the early blockers will still buy it if a copy remains, since its price is fixed and later purchases can only remove backups. Whether a copy remains at her arrival is independent of her own value. We use this observation to compare the numbers of ordinary and high purchases; the formal proof is deferred to Appendix~\ref{app:static-half}.

\paragraph{Welfare assuming Claims~\ref{clm:static-demand} and~\ref{clm:static-competition}.}
Fix the blocker values and first count only buyers who buy their own item. Count each ordinary sale at $a_\ell$ and add its extra value $U\leq\delta$ at the end. In every order there are at most $k$ sales and an expected $\eta$ high buyers per item. This bounds the item's expected welfare by $ka_\ell+\eta(V_\ell-a_\ell)=2ka_\ell$.

On the event in Claim~\ref{clm:static-demand}, consider an offered item $j$ in a layer $\ell<\ell_*$ (any withheld item contributes zero). If $D_{j,b}\leq\eta/2$, at most $\eta$ ordinary buyers purchase in expectation. If $D_{j,b}\geq4k/\eta$, count $a_\ell$ for all $k$ copies and use Claim~\ref{clm:static-competition} for the high buyers' additional value. The respective welfare bounds are
\[
\begin{aligned}
 D_{j,b}\leq\eta/2:\quad&
     \eta a_\ell+\eta V_\ell=(k+2\eta)a_\ell,\\
 D_{j,b}\geq4k/\eta:\quad&
     ka_\ell+(V_\ell-a_\ell)\frac{k\eta}{(1-\eta/N)D_{j,b}}
     \leq k(1+\eta)a_\ell .
\end{aligned}
\]
Both are at most $k(1+2\eta)a_\ell$. The one possible intermediate order contributes at most $2ka_\ell$ and has weight $1/L$ for this item. Thus each earlier layer contributes at most $k(1+2\eta+1/L)$ on average.

In a later layer $\ell>\ell_*$, ordinary buyers can purchase their own item only among the fewer than $\eta m_\ell$ cheap items (since every ordinary value is at most $a_\ell+\delta$). Together they contribute at most $\eta k$, and all high values in the layer have expectation $m_\ell\eta V_\ell=k+\eta$. This gives at most $k(1+2\eta)$ per layer and applies to every layer if $\ell_*$ does not exist. Counting the full $2k$ for layer $\ell_*$ adds at most $k$ to the sum of the layer bounds. Outside the event in Claim~\ref{clm:static-demand}, use the general bound $2dk$ for all layers.

It remains to add the other purchases. All blockers' values have expectation $3\alpha k(L-1)d=3\eta d/8$. The $Nm_\ell$ flexible buyers in layer $\ell<d$ each have backup value at most $2a_{\ell+1}$, so their total backup welfare is at most $2N\sum_{\ell<d}m_\ell a_{\ell+1}=2(d-1)/N$. Together these contributions are at most $\eta kd$. The ordinary buyers' additional values contribute at most $k\delta G$, where $\delta G\leq2\eta$ by geometric growth of the layers. Summing these contributions gives an average welfare of at most
\[
    dk(1+2\eta+1/L)+k+2\eta dk+\eta kd+k\delta G.
\]
Applying Lemma~\ref{lem:positive-prices} with $\gamma=k\delta G$ yields an augmented unit-demand instance and $L$ fixed arrival orders on which every deterministic static pricing mechanism satisfies
\begin{equation}
    \frac1L\sum_{b=0}^{L-1}\welfare(M,\D,K;\pi_b)
    \leq dk(1+2\eta+1/L)+k+2\eta dk+\eta kd+2k\delta G
    \leq dk(1+8\eta),
    \label{eq:static-total}
\end{equation}
where $1/d=1/L\leq\eta$ and $\eta<1/8$. Dividing by \eqref{eq:static-benchmark} gives a ratio at most $(1+8\eta)/(2(1-\eta/4))<1/2+5\eta$. The fixed orders and their probabilities are the same for every mechanism, so averaging over a randomized mechanism's choices proves Theorem~\ref{thm:static-half}. Taking $\eta<\min\{1/8,\epsilon/5\}$ and choosing an order no better than the average gives Proposition~\ref{prop:static-half} for every fixed $k$.
\end{proof}

\section{Supply-Based Pricing with Few Prices}
\label{sec:few-prices}

In this section, we establish improved supply-based pricing guarantees for unit-demand buyers for every fixed number $r\geq 2$ of distinct price levels per item. We first show that just two prices per item achieve a randomized guarantee of $9/13 - o_k(1)$ against the ex-ante optimum (and $2/3 - O(k^{-1/2})$ deterministically), and then extend both mechanisms to any fixed $r\geq 2$, approaching $e/(e+1)$ as $r$ grows.

\paragraph{Tie-breaking.}
The hard instances in Section~\ref{sec:upper-bounds} hold for every tie-breaking rule, even one chosen by the seller. Our algorithms here and in Appendix~\ref{app:randomized-few-prices} work under generic distributions: at every fixed menu, each buyer almost surely has distinct utilities for the individual items and for buying nothing.\footnote{Equivalently, writing \(v_{i0}=0\) for buying nothing, each \(v_{ij}-v_{ih}\) for \(j\ne h\) is atomless. If a price is zero, add the same \(\delta>0\) to every price so that every optimal nonempty bundle is a singleton. For each buyer there are finitely many possible menus, and almost surely her choice at each agrees with the original singleton or empty choice for all sufficiently small \(\delta\). By dominated convergence and finiteness of the arrival orders, the expected welfare loss can be made at most \(\eaopt(\D,K)/k\) in every order when the benchmark is positive.} With adversarial tie-breaking, \textcite{CEFF16} give a three-item unit-demand instance on which every deterministic static price vector admits an order with welfare at most \(2/3\) of optimum. \citet[Section~1.4.2]{TW26} discuss this distinction and extend the static example to any supply by replication. We refer the reader to their paper for a more detailed discussion on tie-breaking.

\subsection{Two price levels}
\label{sec:two-price-guarantee}
Our main positive result for two prices is the following randomized guarantee.

\begin{restatable}{theorem}{twopriceguarantee}
\label{thm:two-price-guarantee}
For unit-demand buyers, there is a randomized supply-based pricing mechanism \(M\) using at most two prices per item such that, for every fixed arrival order \(\pi\),
\[
  \welfare(M,\D,K;\pi)
  \geq \left(\frac{9}{13}-o_k(1)\right)\eaopt(\D,K).
\]
\end{restatable}

The proof of Theorem~\ref{thm:two-price-guarantee} is deferred to Appendix~\ref{app:randomized-two-prices}.
To build toward it, we first construct and analyze a deterministic two-price mechanism guaranteeing $2/3 - O(k^{-1/2})$, which serves as the primary component of our randomized mechanism.

\begin{theorem}
\label{thm:deterministic-two-price}
For unit-demand buyers, there is a deterministic supply-based pricing mechanism \(M\) using at most two prices per item such that, for every fixed arrival order \(\pi\),
\[
  \welfare(M,\D,K;\pi)
  \geq \left(\frac{2}{3}-O(k^{-1/2})\right)\eaopt(\D,K).
\]
\end{theorem}

We describe our intuition on how to choose the two prices $a_j,b_j$ for item $j$. Suppose item \(j\) has \(k_j\) copies and we price the first \(t_j\) copies at \(a_j\) and the remaining copies at \(b_j\). Suppose if under price $a_j$ for each item, the expected number of buyers buying $j$ is $C_j$ and their expected utility is $W_j$. Our social welfare analysis can be separated into three cases. For the intuition, ignore rounding and fluctuations in the number of buyers, and consider how much welfare we can count in each case.

If the first $t_j$ copies does not sold out, we can count the utility from the buyers as $W_j$, as well the revenue from the $C_j$ buyers buying the item at price $a_j$. Thus we count $W_j+a_jC_j$.

If the first $t_j$ copies are sold out while the remaining $k_j-t_j$ does not, we can count the utility from the buyers only toward a price $b_j$ as $W_j-(b_j-a_j)C_j$, and the revenue from buyers buying out the first $t_j$ copies. Thus we count $W_j-(b_j-a_j)C_j+a_jt_j$.

If all copies are sold out, we simply count only the revenue from the seller as $a_jt_j+b_j(k_j-t_j)$.

Note that the optimal social welfare is at most $\sum_j(W_j+a_jk_j)$. At prices $a$, any feasible assignment gives buyers at most $\sum_j W_j$ in expected utility and sells at most $k_j$ copies of item $j$ in expectation. The remaining task is to pick $a_j,b_j,C_j,t_j$ to equalize the social welfare from the above three cases and maximize its ratio comparing to the optimal. The precise parameter choice and analysis is stated below.

\begin{proof}[Proof of Theorem~\ref{thm:deterministic-two-price}]
We first left the choices of $a=(a_1,\ldots,a_m)$ unspecified. At prices $a$, let $L_i$ be buyer $i$'s choice among the items and the option to buy nothing. If $L_i=j$ is still offered at $a_j$ when she arrives, it still maximizes her utility and the mechanism allocates it to her. Write $v_{ij}=v_i(\{j\})$. Then
\[
 C_j=\sum_i\Pr[L_i=j],\qquad
 W_j=\sum_i\E[(v_{ij}-a_j)\one\{L_i=j\}].
\]

We choose the remaining parameters as follows:\footnote{The formulas below apply when $C_j>0$. If $C_j=0$, no buyer chooses $j$ at the base prices, so $W_j=Q_j=0$. The fixed-point formula below then gives $a_j=0$. We set $b_j=0$ as well.}

\begin{equation}\label{eq:few-two-equalize}
 t_j=\frac{k_j+C_j}{2},\qquad b_j=\frac{a_jt_j}{C_j}.
\end{equation}

We choose $a$ so that the welfare $W_j+a_jC_j$ in the first case equals the revenue $a_jt_j+b_j(k_j-t_j)$ in the third. Substituting \eqref{eq:few-two-equalize} gives
\[
 W_j=a_j(t_j-C_j)+b_j(k_j-t_j)
     =\frac{a_j(k_j-C_j)(k_j+3C_j)}{4C_j}.
\]
Thus the three cases give the same welfare bound before rounding and demand fluctuations,
\[
 Q_j=W_j+a_jC_j=\frac{a_j(k_j+C_j)^2}{4C_j}.
\]
And this choice of prices gives
\[
 Q_j-\frac23(W_j+a_jk_j)
 =\frac{a_j(k_j-3C_j)^2}{12C_j}\ge0
 \qquad(C_j>0).
\]
Summing over items and using $\eaopt(\D,K)\le\sum_j(W_j+a_jk_j)$ proves the claimed guarantee.\footnote{In cases where $t_j$ is not an integer, we need to round it down and set the first $\lfloor t_j\rfloor$ copies at price $a_j$. This rounding costs at most $a_j$ at the switch. The base-price count for $j$ has variance at most $C_j$. By averaging, one can fix the auxiliary randomness in the reference choices so that the total expected shortfall in these counts, weighted by $a_j$, is at most $\sum_j a_j\sqrt{C_j}$. This bound on the counts is independent of the order. Since $C_j\le k_j$ and $Q_j\ge a_jk_j$, the combined relative loss is $O(k^{-1/2})$.}

It remains to choose the initial prices, since the buyers' choices themselves depend on these prices. Given tentative counts $c_j$ and values $q_j$, set $x_j=\min\{c_j/k_j,1\}$ and post $p_j=(q_j/k_j)\,4x_j/(1+x_j)^2$. Take all expected counts and values induced by distributions over utility-maximizing choices at these prices. This maps the box $0\le c_j\le n$, $0\le q_j\le\sum_i\E[v_{ij}]$ into itself. Its values are nonempty and convex, and its graph is closed.\footnote{For general priors, take a convergent sequence of tentative counts and values. The corresponding prices converge. The joint laws of a buyer's valuation and chosen item have a weakly convergent subsequence since the valuation has a fixed law and the set of choices is finite. At the limiting prices, the utility gap of each approximating choice is at most twice the largest price change. Since this gap is continuous in the valuation for each choice, the limit law assigns probability one to utility-maximizing choices. Expected counts and values pass to the limit because the valuation marginals are fixed and integrable for a finite benchmark.}
Kakutani's fixed-point theorem~\citep[Corollary to Theorem~1]{Kakutani41} now gives prices $a$ at which the counts and values are precisely $(C_j,Q_j)$ used to set them. If $C_j>k_j$, then $a_j=Q_j/k_j$, whereas every buyer choosing $j$ has value at least $a_j$, so $Q_j\ge a_jC_j$. This forces $Q_j=a_j=0$. On any item with $Q_j=0$ its buyers may choose nothing instead, leaving the prices unchanged. We thus have $C_j\le k_j$, and the price formula at the fixed point gives \eqref{eq:few-two-equalize}.

\end{proof}
\paragraph{From $2/3$ to $9/13$ via randomization.}
The inequality $Q_j \ge \frac{2}{3}(W_j + a_j k_j)$ above is tight only when $C_j = k_j/3$, leaving two-thirds of the copies unsold at the base price $a_j$. To prove Theorem~\ref{thm:two-price-guarantee}, we calibrate the base prices on slightly reduced supplies $\bar k_j = \lfloor (1-\varepsilon)k_j \rfloor$ and randomize globally between two two-price schedules: with probability $10/13$, we run the calibrated schedule of Theorem~\ref{thm:deterministic-two-price}; with probability $3/13$, we instead discount the first $\lfloor \bar k_j - C_j \rfloor$ copies of each item $j$ to attract extra demand while reserving the remaining copies at a slightly raised price. By choosing the reserve prices so that reserved copies almost never stock out, this second strategy preserves nearly all base buyer utility while collecting revenue on the discounted copies, shifting the worst-case demand fraction from $1/3$ to $1/2$ and lifting the guarantee to $9/13 - o_k(1)$. The full proof appears in Appendix~\ref{app:randomized-two-prices}.

\subsection{Extension to \texorpdfstring{\(r\)}{r} price levels}
\label{sec:many-price-guarantee}

Both the randomized and deterministic constructions extend to any fixed number $r\geq2$ of price levels per item. We first state the two guarantees and then describe how the $r$ prices are constructed.

\begin{restatable}{theorem}{manypriceguarantee}
\label{thm:many-price-guarantee}
For every fixed integer \(r\geq2\) and unit-demand buyers, there is a randomized supply-based pricing mechanism \(M_r\) using at most \(r\) prices per item such that, for every fixed arrival order \(\pi\),
\[\welfare(M_r,\D,K;\pi) \geq \left(\frac{1}{1+\left(\frac{r}{r+1}\right)^r}-o_k(1)\right)\eaopt(\D,K).\]
\end{restatable}

\begin{theorem}
\label{thm:deterministic-many-price}
For every integer \(r\geq2\) and unit-demand buyers, there is a deterministic supply-based pricing mechanism \(M_r\) using at most \(r\) prices per item such that, for every fixed arrival order \(\pi\),
\[\welfare(M_r,\D,K;\pi) \geq \left(\frac{1}{1+\bigl((r-1)/r\bigr)^{r-1}} -O\!\left(k^{-1/2}+\frac{r}{k}\right)\right)\eaopt(\D,K).\]
\end{theorem}

The proofs of Theorems~\ref{thm:deterministic-many-price} and~\ref{thm:many-price-guarantee} are given in Appendices~\ref{app:deterministic-many-price} and~\ref{app:randomized-many-prices}, respectively.

\paragraph{Constructing the $r$ price levels.}
For the deterministic mechanism in Theorem~\ref{thm:deterministic-many-price}, the analysis splits into $r+1$ cases for each item $j$. Suppose we offer $t_{j,\ell}$ copies at price $p_{j,\ell}$ for $\ell=1,\ldots,r$, starting at a base price $p_{j,1}=a_j$. If copies remain at $a_j$, we count $W_j+a_jC_j$, just as in Section~\ref{sec:two-price-guarantee}. If the first $\ell-1$ blocks sell out and copies remain at $p_{j,\ell}$, we count
\[W_j-(p_{j,\ell}-a_j)C_j+\sum_{h=1}^{\ell-1}t_{j,h}p_{j,h},\]
combining buyer utility at the higher price with revenue from the earlier blocks. If all $k_j$ copies sell out, we count only the revenue $\sum_{h=1}^r t_{j,h}p_{j,h}$.

Equalizing all $r+1$ cases and maximizing the resulting sellout revenue leads to a geometric schedule. For $C_j>0$, set $\Delta_j=(k_j-C_j)/r$ and $g_j=1+\Delta_j/C_j$. We price the first $C_j+\Delta_j$ copies at $a_j$, and each of the remaining $r-1$ blocks of $\Delta_j$ copies at successive prices $a_jg_j,\ldots,a_jg_j^{r-1}$, choosing the base prices $a$ via Kakutani's fixed-point theorem as before.\footnote{Round the cumulative block sizes down. This only increases revenue at sellout, and a price increase starts less than one copy early, costing at most $p_{j,r}=Q_j/(C_j+\Delta_j)\le rQ_j/k_j$. As in Section~\ref{sec:two-price-guarantee}, demand fluctuations cost at most $\sum_j a_j\sqrt{C_j}\le k^{-1/2}\sum_j Q_j$, since $Q_j=a_jC_jg_j^r\ge a_jk_j$.} Taking $r=2$ recovers Theorem~\ref{thm:deterministic-two-price}.

To obtain Theorem~\ref{thm:many-price-guarantee}, we mix globally between this $r$-price geometric schedule (calibrated on slightly reduced supplies) and the same two-price discount-and-reserve strategy used in Theorem~\ref{thm:two-price-guarantee}; the additional levels are needed only in the first strategy. As $r$ grows, both Theorems~\ref{thm:many-price-guarantee} and~\ref{thm:deterministic-many-price} approach $e/(e+1)$ in the large-supply limit (in particular, taking $r=\lfloor\sqrt{k}\rfloor$ in Theorem~\ref{thm:deterministic-many-price} already approaches $e/(e+1)$ deterministically).
\section{Further Related Work}
\label{sec:related-work}

\paragraph{Single-item multi-unit prophet inequalities.}
\textcite{HKS07} introduced multi-unit prophet inequalities for a single item and showed that a static price approaches the optimal welfare as the supply $k$ grows. For every fixed supply $k$, \textcite{CDL24} set a static price by balancing the expected fraction of copies sold with the probability that at least one copy remains, and \textcite{JMZ23} proved that this choice attains the optimal worst-case guarantee among static thresholds for every $k$.

\paragraph{Few prices and stock-based thresholds.}
Several works study how much a small number of price levels helps for a single resource. When buyers' valuations are known in advance and marginal values for additional units are nonincreasing, \textcite{EFRS20} give a two-price schedule for $k$ identical units that guarantees $5/7 - 1/k$ of the optimal welfare in every arrival order. \textcite{Jiang26} studies a single resource under nonhomogeneous Poisson arrivals with a bounded number of stock-based thresholds.

\paragraph{Single-copy combinatorial prophet inequalities.}
For combinatorial auctions with a single copy of each item ($k=1$), \textcite{EHKS18} obtain a $1-1/e$ guarantee for XOS buyers using time-dependent prices under uniformly random arrival orders. Under arbitrary arrival orders, \textcite{CCFPW24} relate the static-pricing guarantee to bundle size: if every bundle contains a subset of at most $d$ items achieving the same value, static item prices guarantee $1/(d+1)$.

\paragraph{Multi-unit combinatorial settings.}
Much less is known for multi-unit combinatorial auctions. \textcite{CDHKMS17} achieve a ratio of $1-O(\sqrt{\log k/k})$ using static item prices on restricted instances, and \textcite{CMT19} study static bundle prices also under restrictive assumptions on buyers' valuations. For generic request distributions, \textcite{TW26} show how to learn near-optimal static prices in online resource allocation when item capacities are sufficiently large relative to market size---a regime complementary to our static $1/2$ upper bound, which requires $m=\exp(\Omega(k))$.

\section{AI Disclosure}

The author used GPT-6 Astra and Gemini 4 Neon to help develop the results in this paper and polish their presentation.
While working on~\cite{ChawlaDHW25}, the author developed a preliminary idea for a hard instance. The author shared this idea with Astra, which proposed the initial construction in \Cref{sec:two-layer}.
The author then suggested extending the construction to multiple layers. Astra proposed initial versions of the instances in \Cref{sec:layered-supply,sec:static-half}, which the author subsequently simplified.
Astra also proposed initial versions of the algorithms in \Cref{sec:few-prices}. The author simplified the algorithms and their analysis.
The author did not seek to optimize the constants in \Cref{sec:few-prices}. Astra found slight improvements to those constants, so the stated constants are not optimal.
The author takes full responsibility for the correctness and presentation of this paper.

\newpage
\bibliographystyle{ACM-Reference-Format}
\bibliography{arxiv/refs}

\newpage
\appendix
\section{Deferred Proofs for Section~\ref{sec:upper-bounds}}
\label{app:upper-bounds}

\subsection{Reduction to positive prices via initial buyers}
\label{app:initial-buyers}

Under posted pricing, a unit-demand buyer may purchase multiple items if some items are offered at price zero. The following lemma shows that, in all of our upper-bound constructions, we may assume without loss of generality that every offered copy has a strictly positive price---forcing every subsequent unit-demand buyer to purchase at most one item under every tie-breaking rule---at an arbitrarily small additive welfare loss.

\begin{lemma}
\label{lem:positive-prices}
Let $(\D,K)$ be a unit-demand instance with $G$ items, each of supply $k\geq1$, let $\pi_1,\ldots,\pi_L$ be fixed arrival orders with nonnegative weights $w_1,\ldots,w_L$ summing to $1$, and let $\gamma>0$. There exist an augmented unit-demand instance $(\D',K)$ on the same $G$ items (each still of supply $k$) and augmented fixed arrival orders $\pi'_1,\ldots,\pi'_L$ such that $\prophet(\D',K)\geq\prophet(\D,K)$ and the following holds: for every deterministic static (respectively, supply-based) pricing mechanism $M$ on $(\D',K)$ (possibly withholding copies), there is a deterministic static (respectively, supply-based) pricing mechanism $M'$ on $(\D,K)$ (allowing withholding) whose offered prices are all strictly positive, such that
\[
    \sum_{b=1}^{L} w_b\,\welfare(M,\D',K;\pi'_b)
    \leq \sum_{b=1}^{L} w_b\,\welfare(M',\D,K;\pi_b) + \gamma.
\]
\end{lemma}

\begin{proof}
Set $\delta_0=\gamma/(kG)>0$. Form $(\D',K)$ by adding $kG$ deterministic unit-demand \emph{initial buyers}, each of whom values every nonempty bundle at $\delta_0$. For each $b\in[L]$, define $\pi'_b$ by placing these $kG$ initial buyers at the start of the order (in a fixed internal order) followed by $\pi_b$. Because the prophet can ignore the initial buyers, $\prophet(\D',K)\geq\prophet(\D,K)$.

Fix any deterministic static or supply-based pricing mechanism $M$ on $(\D',K)$. Because the initial buyers have deterministic valuations and appear first in every order $\pi'_b$, they face the same menus and make the same purchases in every order, leaving the same remaining copies and prices for the original buyers.

We claim that every copy remaining after the initial buyers has price at least $\delta_0$. Suppose for contradiction that some item still has a copy offered at a price strictly below $\delta_0$ after the initial buyers arrive. Since price schedules are nondecreasing, that item was available at a price strictly below $\delta_0$ throughout the initial buyers' arrivals. Every initial buyer therefore had strictly positive maximum utility and must have purchased at least one copy. With $kG$ initial buyers and at most $kG$ copies in total across all $G$ items, this exhausts every copy, a contradiction.

Thus $M$ induces on the original buyers a deterministic pricing mechanism $M'$ of the same class (static or supply-based, with withholding corresponding to copies sold to the initial buyers) in which every offered copy has price at least $\delta_0>0$. Consequently, any later unit-demand buyer strictly increases her utility by dropping all but one highest-valued item from any bundle of size at least two, so she purchases at most one item under every tie-breaking rule. Finally, since each initial buyer's valuation is at most $\delta_0$, the initial buyers contribute at most $kG\delta_0=\gamma$ to welfare in every order $\pi'_b$, giving $\welfare(M,\D',K;\pi'_b)\leq\welfare(M',\D,K;\pi_b)+\gamma$. Taking the weighted sum over $b\in[L]$ completes the proof.
\end{proof}

\subsection{Proof of Claim~\ref{clm:two-layer-amplification}}
\label{app:two-layer-amplification}

\begin{proof}
Fix a deterministic supply-based mechanism with strictly positive offered prices as in Section~\ref{sec:two-layer}, and retain the notation of the construction. In particular, every buyer purchases at most one item.
Let $q=k^{-k}$ and $\varepsilon=\frac{q}{8k}$.
Choose positive integers $T,N,m$ sufficiently large that
\begin{equation}
\begin{gathered}
    (1+3\varepsilon)^T\geq16k^2,\qquad N\varepsilon^T\geq2,\\
    \frac{N}{m}\leq\varepsilon,\qquad
    \exp\!\left(-\frac{qm}{8k}\right)\leq\frac{1}{2k^2}.
\end{gathered}
\label{eq:two-layer-parameters}
\end{equation}
Recall that $a=1/(2m)$ and $\delta=a/k^3<a$.

For $x\in[0,\delta]$, let $\rho(x)$ be the fraction of backups that are initially unavailable or have initial price at least $2a-x$.
These are precisely the backups whose initial utility to a flexible buyer sampling them is at most $x$.
Since at least $m/k$ backup items are cheap, we have $\rho(x)\leq1-1/k$, and hence $\rho(x)+\varepsilon\leq1$ (as $\varepsilon=q/(8k)<1/k$).

Before any flexible buyer arrives in the quiet order, earlier flexible buyers have bought copies of at most $N$ distinct backups, a fraction at most $N/m\leq\varepsilon$.
A purchase of the main item at price at least $t\in[1,1+\delta]$ requires $1+U\geq t$. Each sampled backup must either be counted in $\rho(1+U-t)$ or have since sold out or risen in price.
Since her samples and $U$ are independent of the preceding history, the expected number of such purchases is at most
\[
    \Lambda(t)\coloneqq\frac{N}{\delta}
        \int_0^{1+\delta-t}\bigl(\rho(x)+\varepsilon\bigr)^T\,dx.
\]
Note that $\Lambda$ is continuous, $\Lambda(1)\geq N\varepsilon^T\geq2$, and $\Lambda(1+\delta)=0$.
Choosing $\tau\in(1,1+\delta)$ with $\Lambda(\tau)=1$ proves part~(i).

In the congested order, all $k$ blockers of an initially cheap backup are active with probability $q$, independently across backups. Since each active blocker values only that backup at $4a>2a$, she purchases a copy whenever one remains at price at most $2a$; thus, when all $k$ blockers are active, the backup either sells out or its price rises strictly above $2a$ before the flexible buyers arrive. Since at least $m/k$ backups are initially cheap, a Chernoff bound gives at least $qm/(2k)=4\varepsilon m$ such backups except with probability at most $\exp(-qm/(8k))\leq1/(2k^2)$.

Condition on blocker values for which this event holds. Fix a flexible buyer's draw $U=u>\tau-1$ and write $x=1+u-\tau>0$. A fraction $\rho(x)$ of backups were initially unavailable or offered utility at most $x$. At least a fraction $4\varepsilon$ of all backups were initially cheap and have all $k$ blockers active. These backups are now unavailable or priced above $2a$; they lie outside the set counted by $\rho(x)$ because their initial utility was at least $a>\delta\geq x$. If all $T$ draws land in their union, every sampled backup offers utility at most $x$ when the buyer arrives, since prices can only rise. Because her utility from the main item at any price strictly below $\tau$ exceeds $x>0$, she strictly prefers the main item to every sampled backup and to buying nothing, and thus purchases it under any tie-breaking rule whenever a copy remains below $\tau$.
Conditional on the blocker realization, this event is independent across flexible buyers, so the expected number of flexible buyers satisfying it is at least
\[
    \frac{N}{\delta}\int_0^{1+\delta-\tau}
        \bigl(\rho(x)+4\varepsilon\bigr)^T\,dx
    \geq (1+3\varepsilon)^T\Lambda(\tau)
    \geq 16k^2,
\]
where the first inequality uses $\rho(x)+\varepsilon\leq1$, which implies $\rho(x)+4\varepsilon\geq(1+3\varepsilon)(\rho(x)+\varepsilon)$.

Since $k\leq8k^2$ for $k\geq2$, a multiplicative Chernoff bound (with deviation $1/2$ from a mean of at least $16k^2$) gives probability at most $\exp(-2k^2)\leq1/(2k^2)$ that fewer than $k$ flexible buyers satisfy this condition. If a copy priced below $\tau$ remained after the flexible buyers, each counted buyer would have bought a copy at a price below $\tau$. With at least $k$ such buyers this exhausts all $k$ copies. Adding the two failure probabilities proves part~(ii).
\end{proof}

\subsection{Proof of Claim~\ref{clm:layered-amplification}}
\label{app:layered-supply}

\begin{proof}
Fix $k\geq1$ and $\eta\in(0,1/8)$, and let $d=\lceil1/\eta\rceil$ as in Section~\ref{sec:layered-supply}.
Let $q=\eta^k$ and $\varepsilon=q/(32d)$, and choose a positive integer $M\geq\max\{2k,8\log(2/\eta)\}$.
Choose positive integers $T,N,m_1,R$, in that order, sufficiently large that
\begin{equation}
\begin{gathered}
    (1+3\varepsilon)^T k\eta\geq M,\qquad N\varepsilon^T\geq2k\eta,\\
    \exp\!\left(-\frac{qm_1}{32}\right)\leq\frac{\eta}{2},\qquad
    R\geq\max\left\{2,\frac{2(N+2k)}{\varepsilon},\frac{2N}{\eta}\right\}.
\end{gathered}
\label{eq:layered-parameters}
\end{equation}
Set $m_\ell=m_1R^{\ell-1}$, $a_\ell=1/m_\ell$, and $\delta=\eta a_d/4$, as in the construction.
We will use
\begin{equation}
    \sum_{u\leq\ell}m_u\leq2m_\ell,
    \qquad
    k\delta\sum_{\ell=1}^d m_\ell\leq\frac{\eta k}{2}.
    \label{eq:layered-size-bounds}
\end{equation}

Fix a deterministic supply-based mechanism with strictly positive offered prices (so every buyer purchases at most one item), an item $j$ in layer $\ell$, and a later layer $t$ with $c_t\geq1/4$.
We use the same count as in Appendix~\ref{app:two-layer-amplification}. Here a backup draw first chooses one of the $d-\ell$ later layers uniformly, then an item uniformly within that layer.
For $x\in[0,\delta]$, let $\rho_\ell(x)$ be the probability that one such draw gives a backup that is initially unavailable or offers utility at most $x$ at its initial price.
Each cheap item in layer $t$ initially offers utility at least $2a_t-(a_t+\delta)=a_t-\delta>\delta$.
A draw hits one of these items with probability at least $1/(4(d-\ell))\geq\varepsilon$, so $\rho_\ell(x)+\varepsilon\leq1$ for every $x\in[0,\delta]$.

In the quiet order, before any flexible buyer of $j$ arrives, only buyers from layers $1,\ldots,\ell$ have arrived.
Since each buys at most one item, together they have bought copies of at most $(N+2k)\sum_{u\leq\ell}m_u\leq2(N+2k)m_\ell$ distinct items.
Every later layer contains at least $Rm_\ell$ items. Even conditional on the layer drawn, the probability of sampling one of the items these buyers purchased is at most $2(N+2k)/R\leq\varepsilon$.
For a buyer to purchase $j$ at a price at least $z\in[a_\ell,a_\ell+\delta]$, her value $a_\ell+U$ must be at least $z$. Each sampled backup must either be counted in $\rho_\ell(a_\ell+U-z)$ or have since sold out or risen in price.
Given the preceding purchases and $U$, her $T$ draws are independent. Averaging over $U$ and summing over the $N$ flexible buyers of $j$ bounds the expected number of such purchases by
\[
    \Lambda_\ell(z)\coloneqq\frac{N}{\delta}
       \int_0^{a_\ell+\delta-z}
           \bigl(\rho_\ell(x)+\varepsilon\bigr)^T\,dx.
\]
Since $\Lambda_\ell$ is continuous, $\Lambda_\ell(a_\ell)\geq N\varepsilon^T\geq2k\eta$, and $\Lambda_\ell(a_\ell+\delta)=0$, we can choose $\tau_j\in(a_\ell,a_\ell+\delta)$ with $\Lambda_\ell(\tau_j)=k\eta$. This proves part~(i).

In the congested order, the blockers of layer $t$ arrive before the flexible buyers of $j$.
Each of the at least $m_t/4$ cheap items in that layer has all $k$ blockers active with probability $q$, independently across items.
By a Chernoff bound, at least $qm_t/8$ cheap items have all $k$ blockers active except with probability at most $\exp(-qm_t/32)\leq\eta/2$.
Each such item sells out or rises to a price strictly above $2a_t$, since each active blocker (of value $4a_t$) buys it whenever a copy remains at price at most $2a_t$.

Fix blocker values in layer $t$ for which this count holds. A single backup draw hits one of these items with probability at least
\[
    \frac{1}{d-\ell}\cdot\frac{qm_t/8}{m_t}
    =\frac{q}{8(d-\ell)}\geq4\varepsilon.
\]
For a flexible buyer with $a_\ell+U>\tau_j$, write $x=a_\ell+U-\tau_j>0$.
These items initially offered utility greater than $\delta\geq x$, so this probability adds to $\rho_\ell(x)$.
If all $T$ draws fall in the union of the two sets, every sampled backup is unavailable or offers utility at most $x$ when the buyer arrives, since prices can only rise. Because her utility from $j$ at any price strictly below $\tau_j$ exceeds $x>0$, she strictly prefers $j$ to every sampled backup and to buying nothing, and therefore purchases $j$ under any tie-breaking rule whenever a copy remains below $\tau_j$.
Conditional on the fixed blocker values, this event is independent across buyers, and for a given $U>\tau_j-a_\ell$ it occurs with probability at least $(\rho_\ell(x)+4\varepsilon)^T$.
Averaging over $U$, the expected number of flexible buyers satisfying this condition is therefore at least
\[
    \frac{N}{\delta}\int_0^{a_\ell+\delta-\tau_j}
       \bigl(\rho_\ell(x)+4\varepsilon\bigr)^T\,dx
    \geq(1+3\varepsilon)^T\Lambda_\ell(\tau_j)
    \geq M.
\]
The first inequality uses $\rho_\ell(x)+\varepsilon\leq1$, as in Appendix~\ref{app:two-layer-amplification}.
Since $M\geq2k$, a multiplicative Chernoff bound gives probability at most $\exp(-M/8)\leq\eta/2$ that fewer than $k$ flexible buyers satisfy this condition.
If a copy priced below $\tau_j$ remained after the flexible buyers, each of these buyers would have bought a copy below $\tau_j$. With at least $k$ such buyers this exhausts the supply.
Adding the two failure probabilities proves part~(ii).
\end{proof}

\subsection{Proofs for the static-pricing bound}
\label{app:static-half}

\subsubsection{Proof of Claim~\ref{clm:static-demand}}

\begin{proof}
Fix $k,\eta$ and take $d,L,\alpha$ as in Section~\ref{sec:static-half}. Put $q=\alpha^k/2$. Choose a positive integer $\lambda$ and then an integer $N$ large enough that
\begin{equation}
    e^{\lambda q\eta/2}>\frac{8k}{\eta^2},\qquad
    N\geq\max\left\{8\lambda,\frac{4k}{\eta}e^{2\lambda d}\right\},
    \qquad (L-1)e^{-q\eta N^2/8}\leq\eta .
    \label{eq:static-parameters}
\end{equation}
Recall that $m_s=N^{2s}$. In particular, $\lambda/m_s\leq1/2$ and each layer has $N^2$ times as many items as the preceding layer.

\paragraph{Counting ordinary demand.}
Fix a deterministic static pricing mechanism with strictly positive offered prices (so every buyer purchases at most one item), the blocker values, and an offered item $j$ in layer $\ell$. For $u>0$ let $n_{s,b}(u)$ count the items in layer $s>\ell$ that are still offered after the early blockers in $\pi_b$ and give backup utility at least $u$. Any such item has price below $2a_s$, so each of its active blockers, of value $3a_s$, buys a copy whenever one remains. Its availability is determined by the number of copies initially offered and the number of early active blockers. Thus $n_{s,b}(u)$ is nonincreasing in $b$.

Each backup in layer $s$ is included independently with probability $\lambda/m_s$. For $u>0$, the probability that none of the sampled backups offers utility at least $u$ is therefore
\begin{equation}
 F_b(u)=\prod_{s>\ell}(1-\lambda/m_s)^{n_{s,b}(u)},
 \qquad D_{j,b}=N\E_U[F_b(a_\ell+U-p_j)],
 \label{eq:static-demand-product}
\end{equation}
where we set $F_b(u)=0$ for $u\leq0$ to include the option of buying nothing. Removing one available backup counted in layer $s$ multiplies this product by $(1-\lambda/m_s)^{-1}$, a factor between $e^{\lambda/m_s}$ and $e^{2\lambda/m_s}$.

By the time a designated buyer of $j$ arrives, at most $N\sum_{h\leq\ell}m_h\leq2Nm_\ell$ earlier flexible buyers could have purchased backups after the early blockers. Each purchases at most one item and hence can exhaust at most one item, while each later layer has at least $N^2m_\ell$ items. Thus, conditional on any preceding purchases, the product increases by a factor at most $\exp(4\lambda/N)<2$. With no later layer the product does not change. The buyer's draws are independent of this history, and conditional on her backup choices the continuous draw $U$ makes equality with her best alternative a probability-zero event. Averaging over $U$ and summing over the $N$ buyers bounds expected ordinary purchases by $2D_{j,b}$.

\paragraph{Comparing successive orders.}
If $p_j\leq a_\ell$, an ordinary buyer strictly prefers $j$ whenever she draws no backups and a copy is offered, almost surely. The probability of drawing no backups is
$\prod_{s>\ell}(1-\lambda/m_s)^{m_s}\geq e^{-2\lambda d}$. Hence $D_{j,b}\geq4k/\eta$ in every order.

Now suppose a layer $\ell_*$ has the cheap fraction specified in the claim. For each cheap item in that layer and each transition from $\pi_b$ to $\pi_{b+1}$, consider the event that its first $b$ blocker batches are inactive and all $k$ blockers in the next batch are active. Its probability is
\[
    (1-\alpha)^{bk}\alpha^k\geq(1-\alpha k(L-1))\alpha^k\geq q.
\]
The item then has a copy left after the early blockers in $\pi_b$ and none in $\pi_{b+1}$, however many copies were initially offered. For each transition these events are independent across the at least $\eta m_{\ell_*}$ cheap items. By a Chernoff bound and a union bound over the $L-1$ transitions, at least $q\eta m_{\ell_*}/2$ such items sell out in every transition except on an event of probability at most $(L-1)e^{-q\eta m_{\ell_*}/8}\leq\eta$.

Fix blocker values for which all transitions satisfy this count, and take $\ell<\ell_*$ and $p_j>a_\ell$. Every positive ordinary surplus $u=a_\ell+U-p_j$ is at most $\delta$. Each of the newly exhausted items, while available, offers utility at least $a_{\ell_*}-\delta>\delta$. Removing its factor from \eqref{eq:static-demand-product} multiplies $F_b(u)$ by at least $e^{\lambda/m_{\ell_*}}$. The other factors can only increase, so
\[
    D_{j,b+1}\geq e^{\lambda q\eta/2}D_{j,b}.
\]
Consequently, once $D_{j,b}>\eta/2$, every later order has demand at least $4k/\eta$. This proves the assertion simultaneously for every offered item in the earlier layers. If no layer is sufficiently cheap, only the ordinary-purchase bound is needed.
\end{proof}

\subsubsection{Proof of Claim~\ref{clm:static-competition}}

\begin{proof}
Fix the blocker values. For each designated buyer $i$ of $j$, let $A_i$ be the event that a copy is still offered just before she arrives. Her own draws are independent of $A_i$. She draws a high value with probability $\eta/N$. With probability
\[
    f=(1-\eta/N)D_{j,b}/N
\]
she draws an ordinary value and strictly prefers $j$ to buying nothing and to all her backups offered just after the early blockers. Such a buyer buys $j$ whenever $A_i$ occurs, since the price of $j$ is fixed and later purchases can only remove backups. These purchases therefore have expected count $f\sum_i\Pr[A_i]\leq k$. A high buyer can buy only when $A_i$ occurs, so the expected number of high purchases is at most
\[
    \frac{\eta}{N}\sum_i\Pr[A_i]
    \leq \frac{k\eta}{(1-\eta/N)D_{j,b}} .\qedhere
\]
\end{proof}
\section{Proofs and Randomized Guarantees for Few Prices}
\label{app:randomized-few-prices}

\subsection{Proof of Theorem~\ref{thm:deterministic-many-price}}
\label{app:deterministic-many-price}

\begin{proof}
Fix an item $j$ with $C_j>0$, and leave the choice of base prices $a$ until the end. To match the first case whenever copies remain at $p_{j,\ell}$, we need the payments from the preceding blocks to equal $C_jp_{j,\ell}$. At the first price increase this requires $t_{j,1}a_j=C_jp_{j,2}$. Each subsequent block adds $t_{j,\ell}p_{j,\ell}$ in payments, so we set
\[
 p_{j,2}=a_j\frac{t_{j,1}}{C_j},\qquad
 p_{j,\ell+1}=p_{j,\ell}\left(1+\frac{t_{j,\ell}}{C_j}\right)
 \quad(2\le\ell<r).
\]
At sellout, the seller has collected $(C_j+t_{j,r})p_{j,r}$. Substituting the prices above, this revenue equals
\[
 a_j C_j\cdot\frac{t_{j,1}}{C_j}
             \prod_{\ell=2}^r\left(1+\frac{t_{j,\ell}}{C_j}\right).
\]
The $r$ factors after $a_jC_j$ sum to $r-1+k_j/C_j$, since the blocks contain $k_j$ copies in total. Their product is largest when they are equal. Writing $\Delta_j=(k_j-C_j)/r$ and $g_j=1+\Delta_j/C_j$, we therefore choose
\[
 t_{j,1}=C_j+\Delta_j,\qquad t_{j,\ell}=\Delta_j\ \ (2\le\ell\le r),
 \qquad p_{j,\ell}=a_jg_j^{\ell-1}\ \ (1\le\ell\le r).
\]
For $r=2$ these are exactly the two prices and the cutoff in \eqref{eq:few-two-equalize}.

It remains to match the sellout revenue $a_jC_jg_j^r$ to the first case. We choose $a$ so that $W_j=a_jC_j(g_j^r-1)$. Then, before rounding and demand fluctuations, all $r+1$ cases give the same welfare bound
\begin{equation}\label{eq:few-base-calibration}
 Q_j=W_j+a_jC_j=a_jC_jg_j^r.
\end{equation}
As in Section~\ref{sec:two-price-guarantee}, the utility comparison holds in expectation: an independent copy of each buyer has the same expected utility at her arrival prices, and can still choose her designated item if it remains available at the final prices.

To compare $Q_j$ with $W_j+a_jk_j$, write $y=(k_j-C_j)/C_j$. Then
\[
 \frac{W_j+a_jk_j}{Q_j}
 =1+\frac{a_j(k_j-C_j)}{Q_j}
 =1+\frac{y}{(1+y/r)^r}.
\]
The fraction is largest at $y=r/(r-1)$, where it equals $((r-1)/r)^{r-1}$. Summing over items gives
\[
 \eaopt(\D,K)\le\sum_j(W_j+a_jk_j)
 \le \left(1+\left(\frac{r-1}{r}\right)^{r-1}\right)\sum_j Q_j .
\]

By the error estimate in Section~\ref{sec:many-price-guarantee}, the expected welfare is at least $(1-k^{-1/2}-r/k)\sum_j Q_j$, which gives the stated guarantee. We may also use a smaller supply to set the prices, keep the last price on any extra copies, and stop counting the item's contribution once that smaller supply sells out. The same estimate then holds with $k$ equal to the minimum supply used to set prices.

For the fixed point in Section~\ref{sec:two-price-guarantee}, replace $4x/(1+x)^2$ by
\[
 A_r(x)=\frac{r^r x^{r-1}}{(1+(r-1)x)^r},
 \qquad a_j=\frac{Q_j}{k_j}A_r(C_j/k_j).
\]
This multiplier gives \eqref{eq:few-base-calibration} and is continuous on $[0,1]$ with $A_r(0)=0$ and $A_r(1)=1$, so the same argument applies.
\end{proof}

\subsection{Proof of Theorem~\ref{thm:two-price-guarantee}}
\label{app:randomized-two-prices}

The deterministic two-price bound of $2/3$ in Section~\ref{sec:two-price-guarantee} is tight only when base-price demand is one third of supply ($C_j=k_j/3$). Mixing with a second two-price strategy---which discounts the first copies to attract extra buyers while reserving the remaining copies for base-price buyers---shifts the worst case to $C_j=k_j/2$ and raises the guarantee to $9/13-o_k(1)$.

\twopriceguarantee*

\begin{proof}
Assume $k$ is sufficiently large that $\varepsilon\coloneqq 4\sqrt{\log(k)/k}\leq 1/2$ (smaller $k$ are absorbed into the $o_k(1)$ term). To create a small buffer of spare copies, we calibrate prices using reduced supplies $\bar k_j=\lfloor(1-\varepsilon)k_j\rfloor$, while still offering all $k_j$ copies of each item. Let $B=\eaopt(\D,\bar K)$ be the ex-ante optimum at the reduced supplies $\bar K=(\bar k_1,\ldots,\bar k_m)$. Scaling any feasible ex-ante allocation for $K$ by $1-\varepsilon-1/k$ shows that
\begin{equation}\label{eq:few-buffer}
 B\geq (1-\varepsilon-1/k)\,\eaopt(\D,K).
\end{equation}

Apply the deterministic two-price calibration of Section~\ref{sec:two-price-guarantee} to the reduced supplies $\bar K$, obtaining base prices $a$, expected base-price demands $C_j\leq\bar k_j$, and welfare bounds $Q_j=W_j+a_jC_j$. Write $A_j=a_j\bar k_j$, $x_j=C_j/\bar k_j\in[0,1]$, and $f(x)=(1+x)^2/(4x)$. For each item with $C_j>0$, the calibration gives $Q_j=A_j f(x_j)$; items with $C_j=0$ have $a_j=W_j=Q_j=0$ and contribute zero to all quantities below. Let $U(a)=\sum_j W_j$ be the total expected buyer utility at base prices $a$, and set
\[
 Q=\sum_j Q_j = U(a)+\sum_j A_j x_j,\qquad
 D=U(a)+\sum_j A_j,\qquad
 H=D-B.
\]
Since $Q\leq B\leq D$, we have $H\geq0$.

Our randomized mechanism chooses globally between two two-price strategies before any buyer arrives:
\begin{itemize}[leftmargin=*,itemsep=3pt]
 \item \emph{Strategy~1 (probability $10/13$):} Run the calibrated two-price schedule of Section~\ref{sec:two-price-guarantee} for $\bar K$, keeping the higher price $b_j$ on the $k_j-\bar k_j$ spare copies. Ignoring welfare after the first $\bar k_j$ copies sell gives expected welfare at least $Q-O(k^{-1/2})Q\geq Q-o_k(B)$.
 \item \emph{Strategy~2 (probability $3/13$):} Offer the first $q_j=\lfloor\bar k_j-C_j\rfloor$ copies of each item $j$ at a discounted price $d_j\leq a_j$ and the remaining $R_j=k_j-q_j$ copies at a reserve price $p_j\geq a_j$. By Lemma~\ref{lem:few-protected} (proved in Appendix~\ref{app:protected-price-proof}), there exist menus $d\leq a\leq p$ guaranteeing expected welfare at least $S-2H-o_k(B)$ in every fixed arrival order, where
 \[
  S=U(a)+\sum_j A_j\min\{1-x_j,2/3\}.
 \]
\end{itemize}

We compare both guarantees to $D$ item by item. Expressing $U(a)=Q-\sum_j A_j x_j=\sum_j A_j(f(x_j)-x_j)$, each item $j$ with $C_j>0$ contributes $f(x_j)+1-x_j$ to $D$ (in units of $A_j$), $1-x_j$ to $D-Q$, and $\max\{x_j,1/3\}$ to $D-S$. We claim that $10Q+3S\geq 9D$, equivalently $10(D-Q)+3(D-S)\leq 4D$. Dropping the index $j$, the per-item slack in units of $A_j$ is
\[
 4\bigl[f(x)+1-x\bigr]-10(1-x)-3\max\{x,1/3\}
 =\frac{1}{x}+7x-4-3\max\{x,1/3\}.
\]
For $x\geq 1/3$, this equals $\frac{1}{x}+4x-4=\frac{(2x-1)^2}{x}\geq0$ (tight at $x=1/2$). For $x\leq 1/3$, it equals $\frac{1}{x}+7x-5\geq 2\sqrt{7}-5>0$ by AM--GM. Summing over items gives $10Q+3S\geq 9D$.

Consequently, the mixed mechanism achieves expected welfare at least
\[
 \frac{10}{13}Q+\frac{3}{13}(S-2H)-o_k(B)
 \geq \frac{9}{13}D-\frac{6}{13}H-o_k(B)
 =\frac{9}{13}B+\frac{3}{13}H-o_k(B)
 \geq \left(\frac{9}{13}-o_k(1)\right)B,
\]
using $D=B+H$ and $H\geq0$. Applying \eqref{eq:few-buffer} completes the proof.
\end{proof}

\subsection{Proof of Theorem~\ref{thm:many-price-guarantee}}
\label{app:randomized-many-prices}

For general $r\geq2$, we keep the same two-price discount-and-reserve mechanism as Strategy~2 and upgrade Strategy~1 to the $r$-price geometric schedule of Appendix~\ref{app:deterministic-many-price}.

\manypriceguarantee*

\begin{proof}
Use the same reduced supplies $\bar K$, buffer bound \eqref{eq:few-buffer}, and notation $B,A_j,x_j,U(a),Q,D,H$ as in Appendix~\ref{app:randomized-two-prices}, now calibrated via the $r$-price schedule of Appendix~\ref{app:deterministic-many-price}. By \eqref{eq:few-base-calibration}, every item with $C_j>0$ satisfies $Q_j=A_j f_r(x_j)$, where
\[
 f_r(x)\coloneqq x\left(1+\frac{1-x}{rx}\right)^r
 =\frac{\bigl(1+(r-1)x\bigr)^r}{r^r x^{r-1}}.
\]
Expanding $(1+(r-1)x)^r$ binomially writes $f_r(x)$ as a positive linear combination of $x^{t-r+1}$ for $t=0,\ldots,r$; since every exponent is either linear, constant, or negative, $f_r$ is convex on $(0,\infty)$. Differentiating $\log f_r(x)$ gives
\begin{equation}\label{eq:fr-derivative}
 \frac{f'_r(x)}{f_r(x)}
 =\frac{r(r-1)}{1+(r-1)x}-\frac{r-1}{x}
 =-\frac{(r-1)(1-x)}{x\bigl(1+(r-1)x\bigr)}\leq0
 \qquad\text{for }x\in(0,1],
\end{equation}
so $f_r$ is decreasing on $(0,1]$.

Set $T_r=(1+1/r)^r$, $c_r=T_r/(1+T_r)=\bigl[1+(r/(r+1))^r\bigr]^{-1}$, and $\beta_r=c_r/(r+1)$. Our mechanism runs Strategy~2 (from Lemma~\ref{lem:few-protected}, using the new reference data) with probability $\beta_r$, and Strategy~1 (the $r$-price geometric schedule for $\bar K$, keeping the last price on all spare copies) with probability $1-\beta_r$. By Appendix~\ref{app:deterministic-many-price} and Lemma~\ref{lem:few-protected}, these achieve expected welfare at least $Q-o_k(B)$ and $S-2H-o_k(B)$, respectively.

We claim that $(1-\beta_r)Q+\beta_r S\geq c_r D$, equivalently $(1-\beta_r)(D-Q)+\beta_r(D-S)\leq (1-c_r)D$. Since $1-c_r=1/(1+T_r)$ and $(1+T_r)\beta_r=T_r/(r+1)$, multiplying the per-item slack (in units of $A_j$) by $1+T_r$ yields
\[
 g(x)\coloneqq f_r(x)+1-x-\left(1+\frac{rT_r}{r+1}\right)(1-x)-\frac{T_r}{r+1}\max\{x,1/3\}
 =f_r(x)-\frac{rT_r}{r+1}(1-x)-\frac{T_r}{r+1}\max\{x,1/3\}.
\]
We verify $g(x)\geq0$ on $(0,1]$ in two ranges:
\begin{itemize}[leftmargin=*,itemsep=3pt]
 \item \emph{For $x\geq 1/3$:} We have $g(x)=f_r(x)-\frac{T_r}{r+1}\bigl(r-(r-1)x\bigr)$. Direct substitution gives $f_r(1/2)=T_r/2$ and hence $g(1/2)=0$, while \eqref{eq:fr-derivative} gives $f'_r(1/2)=-\frac{r-1}{r+1}T_r$ and hence $g'(1/2)=0$. Since $f_r$ is convex, $g$ is convex on $[1/3,1]$ and therefore $g(x)\geq g(1/2)=0$.
 \item \emph{For $x\leq 1/3$:} Here $g'(x)=f'_r(x)+\frac{r}{r+1}T_r$. By convexity and monotonicity of $f_r$, for every $x\leq 1/3$ we have $f'_r(x)\leq f'_r(1/3)$ and $f_r(1/3)\geq f_r(1/2)=T_r/2$. Using \eqref{eq:fr-derivative} at $x=1/3$,
 \[
  f'_r(1/3)=-\frac{6(r-1)}{r+2}f_r(1/3)
  \leq-\frac{3(r-1)}{r+2}T_r
  <-\frac{r}{r+1}T_r,
 \]
 where the strict inequality holds because $3(r-1)(r+1)-r(r+2)=2r^2-2r-3>0$ for all $r\geq2$. Thus $g$ is strictly decreasing on $(0,1/3]$, so $g(x)\geq g(1/3)\geq0$.
\end{itemize}

Summing over items yields $(1-\beta_r)Q+\beta_r S\geq c_r D$. Since $r\geq2$ implies $2\beta_r=\frac{2}{r+1}c_r<c_r$, the mixed mechanism achieves expected welfare at least
\[
 c_r D-2\beta_r H-o_k(B)
 =c_r B+(c_r-2\beta_r)H-o_k(B)
 \geq \bigl(c_r-o_k(1)\bigr)B.
\]
Applying \eqref{eq:few-buffer} finishes the proof.
\end{proof}

\subsection{Analysis of the discount-and-reserve strategy}
\label{app:protected-price-proof}

We now construct the two-price menu $d\leq a\leq p$ for Strategy~2 and prove its welfare guarantee. Under Strategy~2, each item $j$ has $q_j=\lfloor\bar k_j-C_j\rfloor$ discounted copies at price $d_j$ and $R_j=k_j-q_j\geq C_j+\varepsilon k_j$ reserved copies at price $p_j$.

To guarantee that reserved copies rarely sell out, we must account for buyers who spill over to a backup item $j$ after all items they prefer to $j$ have stocked out. If each item stocks out with probability at most $\rho$, an item in position $\ell$ of a buyer's preference list should be reached only with probability about $\rho^{\ell-1}$. Lemma~\ref{lem:few-reserved-copies} formalizes this exponential-decay bound, and Lemma~\ref{lem:few-protected} then sets $d$ and $p$ by minimizing two simple convex objectives whose gradients equal expected demands and rank-weighted loads.

\begin{lemma}\label{lem:few-reserved-copies}
Fix an arrival order. Each buyer has an ordered list of distinct items, independent of the history before she arrives. Upon arrival, she either skips or takes the first item on her list with a reserved copy left (her decision to skip may depend on her own information and the history). Let $R_j\leq k_j$ be the initial reserve of item $j$, and let $\Lambda_j$ be its expected rank-weighted load across all buyers, where an appearance in position $\ell$ contributes weight $\rho^{\ell-1}$ for a parameter $\rho\in(0,1)$. Suppose $\Lambda_j\leq R_j-\eta k_j$ for every $j$, with $\eta\in(0,1]$. If $e^{-\eta^2 k/2}\leq\rho$, then each item $j$ exhausts its reserve with probability at most $e^{-\eta^2 k_j/2}$.
\end{lemma}

\begin{proof}
For any subset $S\subseteq[m]$, let $X_t(S)$ be the number of reserved copies sold from $S$ to the first $t$ buyers, let $L_t(S)$ be the sum of the first $t$ buyers' expected rank weights on $S$, and write $R(S)=\sum_{j\in S}R_j$. Set $z=e^\eta-1$. For every $j$ and $t$, using $L_t(\{j\})\leq\Lambda_j\leq R_j-\eta k_j$ and $R_j\leq k_j$ (noting $e^\eta-1-\eta\geq0$),
\begin{equation}\label{eq:mgf-decay}
 zL_t(\{j\})-\eta R_j
 \leq (e^\eta-1)(R_j-\eta k_j)-\eta R_j
 \leq \bigl((1-\eta)e^\eta-1\bigr)k_j
 \leq -\frac{\eta^2 k_j}{2}
 \leq \log\rho,
\end{equation}
where the third inequality uses $1-(1-\eta)e^\eta=\int_0^\eta u e^u\,du\geq\eta^2/2$ and the last uses $k_j\geq k$ and $e^{-\eta^2 k/2}\leq\rho$.

We prove by induction on $t$ that $\E[e^{\eta X_t(S)}]\leq e^{zL_t(S)}$ for every $S\subseteq[m]$. The base case $t=0$ is trivial. For the inductive step, condition on buyer $t$'s preference list. If the list intersects $S$, let $T$ be the prefix of items appearing strictly before the list's first item in $S$; otherwise buyer $t$ cannot take from $S$. For buyer $t$ to take a reserved copy from $S$, every item in $T$ must already have exhausted its reserve, so $X_{t-1}(T)=R(T)$. Since $S$ and $T$ are disjoint, $\one\{X_{t-1}(T)=R(T)\}\leq e^{\eta(X_{t-1}(T)-R(T))}$, and buyer $t$'s list is independent of $X_{t-1}$, the induction hypothesis on $S\cup T$ yields
\[
\begin{aligned}
 \E\!\left[e^{\eta X_{t-1}(S)}\one\{\text{buyer $t$ takes from $S$}\}\,\middle|\,\text{list}\right]
 &\leq e^{-\eta R(T)}\E\!\left[e^{\eta X_{t-1}(S\cup T)}\right] \\
 &\leq e^{-\eta R(T)}e^{z L_{t-1}(S\cup T)}
 =e^{zL_{t-1}(S)}\prod_{j\in T}e^{zL_{t-1}(\{j\})-\eta R_j}
 \leq e^{zL_{t-1}(S)}\rho^{|T|},
\end{aligned}
\]
where the last step applies \eqref{eq:mgf-decay}. Here $\rho^{|T|}$ is the rank weight of the list's first item in $S$, which is at most the list's total rank weight on $S$. Taking the expectation over buyer $t$'s list bounds the left-hand side by $e^{zL_{t-1}(S)}\bigl(L_t(S)-L_{t-1}(S)\bigr)$. Since a purchase from $S$ multiplies $e^{\eta X_{t-1}(S)}$ by $e^\eta=1+z$, we obtain
\[
 \E\!\left[e^{\eta X_t(S)}\right]
 =\E\!\left[e^{\eta X_{t-1}(S)}\right]
  +z\,\E\!\left[e^{\eta X_{t-1}(S)}\one\{\text{buyer $t$ takes from $S$}\}\right]
 \leq e^{zL_{t-1}(S)}\bigl(1+z[L_t(S)-L_{t-1}(S)]\bigr)
 \leq e^{zL_t(S)},
\]
closing the induction. Finally, applying Markov's inequality and \eqref{eq:mgf-decay} with $S=\{j\}$ gives
\[
 \Pr\bigl[X_n(\{j\})=R_j\bigr]
 \leq e^{-\eta R_j}\E\!\left[e^{\eta X_n(\{j\})}\right]
 \leq e^{zL_n(\{j\})-\eta R_j}
 \leq e^{-\eta^2 k_j/2}.\qedhere
\]
\end{proof}

\begin{lemma}\label{lem:few-protected}
Let $\varepsilon=4\sqrt{\log(k)/k}\leq 1/2$. For any fixed $r\geq2$ and the corresponding calibrated reference prices $a$ and demands $C_j\leq\bar k_j$ at reduced supplies $\bar K$, there exist menus $d\leq a\leq p$ such that Strategy~2 achieves expected welfare at least $S-2H-O(k^{-1/2})B$ in every fixed arrival order.
\end{lemma}

\begin{proof}
Set $q_j=\lfloor\bar k_j-C_j\rfloor$ and $R_j=k_j-q_j$, so $C_j+\varepsilon k_j\leq R_j\leq k_j$. For any menu $z\in\R_{\geq0}^m$, write $u_i(v_i,z)=\max\{0,\max_j(v_{ij}-z_j)\}$ and $U(z)=\sum_i\E[u_i(v_i,z)]$. Because the distributions are generic, at any fixed menu $z$ buyer $i$'s utilities for all items and for buying nothing are almost surely strictly distinct, so her optimal choice is almost surely unique. Since $u_i(v_i,\cdot)$ is $1$-Lipschitz in each coordinate, dominated convergence shows that $U$ is differentiable everywhere with $\frac{\partial U}{\partial z_j}(z)=-\sum_i\Pr[\text{buyer $i$ chooses $j$ at $z$}]$.

\paragraph{1. Choosing the discounted prices $d\leq a$.}
Let
\[
 d\in\argmin_{0\leq z\leq a}\left(U(z)+\frac{2}{3}\sum_j \bar k_j z_j\right).
\]
Let $Y_i$ be buyer $i$'s (a.s.\ unique) utility-maximizing choice at menu $d$, and let $N_j=\sum_i\one\{Y_i=j\}$. Whenever $d_j>0$, first-order optimality with respect to decreasing $z_j$ gives $\frac{\partial U}{\partial z_j}(d)+\frac{2}{3}\bar k_j\leq0$, i.e.,
\[
 \E[N_j]\geq\frac{2}{3}\bar k_j\qquad\text{whenever }d_j>0.
\]
Comparing the objective at $d$ and $a$ gives $U(d)-U(a)\leq\frac{2}{3}\sum_j \bar k_j(a_j-d_j)$. Adding this to the standard dual bound $B\leq U(d)+\sum_j \bar k_j d_j$ and substituting $B=D-H=U(a)+\sum_j \bar k_j a_j-H$ yields $\frac{1}{3}\sum_j \bar k_j(a_j-d_j)\leq H$, or
\begin{equation}\label{eq:few-markdown}
 \sum_j \bar k_j(a_j-d_j)\leq 3H.
\end{equation}

\paragraph{2. Choosing the reserve prices $p\geq a$.}
Set $\xi=\eta=\varepsilon/2=2\sqrt{\log(k)/k}$, $\rho=k^{-2}$, and $c_j=C_j+\xi k_j$. For each buyer $i$ and menu $z$, define
\[
 \psi_{\rho,i}(v_i,z)=\max_{(j_1,\ldots,j_s)}\sum_{\ell=1}^s \rho^{\ell-1}(v_{ij_\ell}-z_{j_\ell}),
 \qquad
 \Psi_\rho(z)=\sum_i\E\bigl[\psi_{\rho,i}(v_i,z)\bigr],
\]
where the maximum ranges over all ordered lists of distinct items (including the empty list). Because $1>\rho>\rho^2>\cdots>0$, the maximizing list for buyer $i$ at $z$ consists of all items with $v_{ij}-z_j>0$ sorted in decreasing order of $v_{ij}-z_j$; under generic distributions, this list is almost surely unique. As $\psi_{\rho,i}(v_i,\cdot)$ is $1$-Lipschitz in each coordinate, dominated convergence gives $\frac{\partial \Psi_\rho}{\partial z_j}(z)=-\Lambda_j(z)$, where $\Lambda_j(z)$ is the expected rank-weighted load of $j$ across buyers' preference lists at $z$.

Choose
\[
 p\in\argmin_{z\geq a}\left(\Psi_\rho(z)+\sum_j c_j z_j\right),
\]
which exists because $\Psi_\rho\geq0$ is continuous and $c_j>0$. First-order optimality with respect to increasing $z_j$ yields $\Lambda_j(p)\leq c_j\leq R_j-\eta k_j$ for every $j$. Furthermore, every list satisfies $u_i(v_i,z)\leq\psi_{\rho,i}(v_i,z)\leq u_i(v_i,z)/(1-\rho)$, so $U(z)\leq\Psi_\rho(z)\leq U(z)/(1-\rho)$. Comparing $p$ with $a$ and using $U(p)\geq U(a)-\sum_j C_j(p_j-a_j)$ (as buyers can still choose their base choices $L_i$ at $p$) gives
\begin{equation}\label{eq:few-upper-movement}
 \xi\sum_j k_j(p_j-a_j)
 =\sum_j(c_j-C_j)(p_j-a_j)
 \leq \Psi_\rho(a)-\Psi_\rho(p)+\sum_j C_j(p_j-a_j)
 \leq \frac{\rho}{1-\rho}U(a).
\end{equation}

\paragraph{3. Bounding discounted revenue and buyer utility.}
We lower bound the mechanism's welfare by summing the revenue from discounted copies and the total buyer utility (dropping nonnegative revenue from reserved copies).

\emph{Discounted revenue.}
Whenever $Y_i=j$ and a discounted copy of $j$ remains upon buyer $i$'s arrival, item $j$ is offered at $d_j$ while every other item has current price at least $d_h$; thus $j$ still uniquely maximizes buyer $i$'s utility and she purchases a discounted copy of $j$. Hence at least $\min\{q_j,N_j\}$ discounted copies of $j$ sell. Set $m_j=\min\{q_j,\frac{2}{3}\bar k_j\}$. Whenever $d_j>0$, we have $\E[N_j]\geq m_j$; thinning each indicator $\one\{Y_i=j\}$ independently with probability $m_j/\E[N_j]$ produces a sum $\tilde N_j\leq N_j$ of independent Bernoullis with mean $m_j$ and variance at most $m_j$, so
\[
 \E\bigl[\min\{q_j,N_j\}\bigr]
 \geq \E\bigl[\min\{m_j,\tilde N_j\}\bigr]
 =m_j-\E\bigl[(m_j-\tilde N_j)_+\bigr]
 \geq m_j-\sqrt{\mathrm{Var}(\tilde N_j)}
 \geq m_j-\sqrt{m_j}.
\]
Since $m_j\geq \bar k_j\min\{1-x_j,2/3\}-1$ and $d_j\leq a_j$, summing $d_j\,\E[\min\{q_j,N_j\}]$ over all items and applying \eqref{eq:few-markdown} gives expected discounted revenue at least
\[
\begin{aligned}
 \sum_j d_j\bar k_j\min\{1-x_j,2/3\}-\sum_j a_j\bigl(1+\sqrt{\bar k_j}\bigr)
 &\geq \sum_j A_j\min\{1-x_j,2/3\}-\frac{2}{3}\sum_j \bar k_j(a_j-d_j)-O(k^{-1/2})\sum_j A_j \\
 &\geq \sum_j A_j\min\{1-x_j,2/3\}-2H-O(k^{-1/2})Q,
\end{aligned}
\]
where the last step uses $\sum_j A_j\leq Q$ (since $f_r(x)\geq x(1+(1-x)/x)=1$ by Bernoulli's inequality).

\emph{Buyer utility.}
Consider buyer $i$'s preference list at menu $p$. If she purchases a reserved copy of some item $j$ (at price $p_j$), every item $h$ ranked strictly ahead of $j$ on that list must already be sold out---otherwise $h$ would still be available at price $d_h$ or $p_h$, giving strictly higher utility than $j$ at $p_j$. Thus every buyer who purchases a reserved copy takes the first item on her preference list with a reserved copy left, while buyers who buy a discounted copy or buy nothing skip the reserves. Since $\Lambda_j(p)\leq R_j-\eta k_j$ and $e^{-\eta^2 k/2}=k^{-2}=\rho$, Lemma~\ref{lem:few-reserved-copies} implies that each item $j$ stocks out with probability at most $k^{-2}$.

Let $E_{ij}$ be the event that item $j$ has not stocked out before buyer $i$ arrives; $E_{ij}$ is independent of buyer $i$'s valuation and base choice $L_i$, and $\Pr[E_{ij}]\geq 1-k^{-2}$. Whenever $L_i=j$ and $E_{ij}$ holds, item $j$ is offered at price at most $p_j$, so buyer $i$ obtains utility at least $\max\{0,v_{ij}-p_j\}\geq (v_{ij}-a_j)-(p_j-a_j)$. Summing expectations over all buyers and using \eqref{eq:few-upper-movement} (with $C_j\leq k_j$, $\rho=k^{-2}\leq 1/2$, and $\xi=2\sqrt{\log(k)/k}$) gives total expected buyer utility at least
\[
 (1-k^{-2})U(a)-\sum_j C_j(p_j-a_j)
 \geq \left(1-k^{-2}-\frac{2\rho}{\xi}\right)U(a)
 =U(a)-O(k^{-3/2})Q.
\]
Adding the discounted revenue and buyer utility bounds gives expected welfare at least $S-2H-O(k^{-1/2})Q\geq S-2H-O(k^{-1/2})B$, since $Q\leq B$.
\end{proof}

\end{document}